\documentclass[showpacs,preprint,superscriptaddress,amsmath,amssymb]{revtex4-1}
\usepackage{graphicx}
\usepackage{dcolumn}
\usepackage{bm}
\usepackage{epsfig}
\usepackage{subfigure}
\usepackage{graphics}
\usepackage{amssymb}
\usepackage{amsthm}
\usepackage{amsmath}
\usepackage{array}
\usepackage{color}
\usepackage{booktabs}

\begin{document}

\preprint{APS/123-QED}

\title{Numerical study of three-dimensional single-mode Rayleigh-Taylor instability in turbulent mixing stage}

\author{Bin Liu}
 \affiliation{Department of Physics, Hangzhou Dianzi University, Hangzhou 310018, China}

\author{Chunhua Zhang}
\affiliation{Department of Mechanics and Aerospace Engineering, Southern University of Science and Technology, Shenzhen 518055, China}

  \author{Qin Lou}
 \affiliation{School of Energy and Power Engineering, University of Shanghai for Science and Technology, Shanghai 200093, China}

\author{Hong Liang}
\email[Corresponding author]{: lianghongstefanie@163.com}
\affiliation{Department of Physics, Hangzhou Dianzi University, Hangzhou 310018, China}

\date{\today}

\begin{abstract}
Rayleigh-Taylor instability (RTI) as a multi-scale, strongly nonlinear physical phenomenon which plays an important role in the engineering applications and scientific research. In this paper, the mesoscopic lattice Boltzmann method is used to numerically study the late-time evolutional mechanism of three-dimensional (3D) single-mode RTI and the influences of extensive dimensionless Reynolds number and Atwood number on phase interfacial dynamics, spike and bubble growth are investigated in details. For a high Reynolds number, it is reported that the development of 3D single-mode RTI would undergo four different stages: linear growth stage, saturated velocity growth stage, reacceleration stage and turbulent mixing stage. A series of complex interfacial structures with large topological changes can be observed at the turbulent mixing stage, which always preserve the symmetries with respect to the middle axis at a low Atwood number, and the lines of symmetry within spike and bubble are broken as the Atwood number is increased. Five statistical methods for computing the spike and bubble growth rates are then analyzed to reveal the growth law of 3D single-mode RTI in turbulent mixing stage. It is found that the spike late-time growth rate shows an overall increase with the Atwood number, while the bubble growth rate first decreases slightly with the Atwood number and then approaches a constant of around 0.1.  When the Reynolds number decreases, the later stages cannot be reached gradually and the evolution of phase interface presents a laminar flow state.

\end{abstract}
\maketitle

\section{Introduction}
Rayleigh-Taylor instability (RTI) is a phenomenon caused by small perturbations at the interface of two varying-density fluids under the action of acceleration. This  instability phenomenon can be widely encountered in natural phenomena such as the formation of cirrus clouds, supernova explosion~\cite{Burrows} and the formation of underground salt domes. Also, it plays a vital role in turbulent mixing science~\cite{Chertkov} and many engineering applications including inertial confinement nuclear fusion~\cite{Betti}, meteorology and ocean kinematics. Since the pioneering work by Rayleigh and Taylor~\cite{Rayleigh}, a great number of theoretical, numerical and experimental studies have been dedicated to this classical interfacial instability, which has also been the subject of several recent reviews~\cite{Zhou1, Zhou2, Boffetta, Livescu}.

Based on the type of initial perturbation, the RTI can be divided into the single-mode and multi-mode~\cite{Tavares, Adkins} examples. According to the latest knowledge~\cite{Ramaprabhu1, Wei}, it is well acknowledged that the development of single-mode RTI can be divided into four different stages: linear growth stage, saturated velocity stage, reacceleration stage and chaotic or turbulent mixing stage. Rayleigh and Taylor~\cite{Rayleigh} were the first to present linear stability theory indicating the disturbance amplitude of inviscid fluids increases exponentially with time. Later, the experiment by Lewis~\cite{Lewis} revealed that the linear stability theory was effective to describe the growth of the perturbation until its amplitude reaching $0.4\lambda$, where $\lambda$ is the initial wavelength. Bellman and Pennington~\cite{Bellman} analyzed the influence of fluid viscosity and surface tension on instability linear growth and formulated an implicit relation for linear growth rate, which was further extended to an explicit one~\cite{Menikoff} by simplifying the high-order term of Bellman's model~\cite{Bellman}. For the saturated velocity stage, the single-mode RTI would evolve with approximately constant velocity and several analytical models have been proposed for depicting spike and bubble quasisteady velocities. In 1955, Layzer~\cite{Layzer} proposed the first potential flow model for RTI saturated stage applied only to the fluid-vacuum interfaces. On the basis of the Layzer model~\cite{Layzer}, Goncharov~\cite{Goncharov} used a different form of velocity potential and then extended the Layzer-type model to the system of arbitrary Atwood numbers. Enlightened by the above work~\cite{Goncharov}, some modified potential flow models are proposed in succession by incorporating the effects of fluid viscosity, surface tension~\cite{Sohn} and the vortices~\cite{Betti}. A detailed experiment of two-dimensional (2D) single-mode RTI~\cite{Waddell} also showed that the average of the spike and bubble velocities approaches to a constant in the following stage of linear growth.

Following the saturated velocity stage, the so-called reacceleration stage characterizing with increasing velocity could be observed, when the flow Reynolds number is sufficiently high. This stage was first found in a long-time 2D simulation of Glimm {\it{et al.}}~\cite{Glimm} using the front tracking method. They reported that the spike velocity has a significant increase after a time period of plateau. The reacceleration stage was also verified in the later three-dimensional (3D) simulations by Ramaprabhu {\it{et al.}}~\cite{Ramaprabhu2}, who discovered the reacceleration process occurring at the tip of the bubble and ascribed this behavior to the formed Kelvin-Helmholtz vortices on the bubble-spike interface. Afterwards, it was further confirmed in a 3D experiment of single-mode RTI~\cite{Wilkison}, where the bubble and spike tips were found to be accelerated such that their evolutional velocities have exceeded than the predictions of the potential flow theory~\cite{Goncharov}. Bian~\cite{Bian} numerically studied the effect of vorticity on the reacceleration stage of 3D compressible single-mode RTI under different Reynolds numbers and Atwood numbers, and showed a clear correlation between the reacceleration process and vortices inside the bubble. The reacceleration stage cannot last indefinitely followed by the fourth stage named as the chaotic or turbulent mixing stage. Ramaprabhu {\it{et al.}}~\cite{Ramaprabhu1} conducted a late-time simulation of 3D single-mode RTI and primitively observed a sequence of events at high Reynolds numbers that can be summarized in four stages. At the chaotic mixing stage, they reported that the late-time instability was significantly strengthened resulting in the turbulent mixing of fluids and the bubble and spike velocities experience an decrease with time. This was in contradiction with the direct numerical simulation result of single-mode RTI~\cite{Wei} that the late-time bubble acceleration fluctuated with time showing a mean quadric growth. Recently, Liang {\it{et al.}} used an improved lattice Boltzmann model to investigate 2D single-mode RTI with extensive Reynolds numbers and Atwood numbers~\cite{Liang1, Liang2}, and also 3D example~\cite{Liang3} within a low-Atwood-number fluid system. The late-time quadratic growth phenomenon at a high Reynolds number was also observed and the quantitative description of the 2D spike and bubble growth rates was further provided. More recently, Hu {\it{et al.}}~\cite{Hu} simulated the late-time dynamics of 2D single-mode RTI and found that at a medium Reynolds number, the flow will not enter into the chaotic mixing stage instead of a new deceleration-acceleration stage, in which the bubble velocity is decelerated and accelerated repeatedly.

Although several efforts~\cite{Ramaprabhu1, Wei, Liang1, Liang3, Hu} have been made to study the late-time single-mode RTI, the understanding of this instability in the turbulent mixing stage has not been fully addressed. Previously, we have presented a quantitative study on the late-time dynamics of 2D single-mode RTI with various flow parameters~\cite{Liang1}. As a continuous work, in this paper we intend to investigate the late-time growth of 3D single-mode RTI and the quantitative data of the growth rate to determine the nature of the turbulent mixing stage is also provided. The rest of this article is organized as follows. In Sec.~\ref{sec:method}, we will give an introduction of numerical methodology. Sec~\ref{sec:Results} describes the long-time evolution of 3D single-mode instability with different Reynolds numbers and Atwood numbers. Finally, we conduct a summary in Sec.~\ref{sec:summary}.

\section{Numerical methodology}~\label{sec:method}
The lattice Boltzmann method~\cite{Guo} has received great success in simulating complex flows and in particular to multiphase flows. Several different types of multi-phase multi-component lattice Boltzmann models have been proposed~\cite{Wang}, including the color-gradient model, the pseudo-potential model, the free-energy model, and the phase-field based model, among which the last type has shown great potential in solving variable-density interfacial instability~\cite{Liang1, Liang3, Liang4}. The mathematical formulations of two-phase fluid system in the phase field framework consist of the Cahn-Hilliard equation and the incompressible Navier-Stokes equations~\cite{Jacqmin}
\begin{equation}
{{\partial \phi } \over {\partial t}} + \nabla  \cdot (\phi {\bf{u}})
= \nabla  \cdot {M}(\nabla {\mu} ),
\end{equation}
\begin{subequations}
\begin{equation}
\nabla  \cdot {\bf{u}} = 0,
\end{equation}
\begin{equation}
\rho ({{\partial {\bf{u}}} \over{\partial t}} + {\bf{u}}\cdot\nabla {\bf{u}} )=  - \nabla p +\nabla  \cdot \left[ {\nu\rho(\nabla {\bf{u}} + \nabla{{\bf{u}}^T})} \right] + {{\bf{F}}_s} + {\bf{G}},
\end{equation}
\end{subequations}
where $\phi$ is the order parameter for distinguishing phase interface, $\mu$ is the chemical potential, $M$ is the mobility, ${\bf{u}}$ is the fluid velocity, $\rho$ is the fluid density, $\nu$ represents the kinematic viscosity, ${p}$ is the hydrodynamic pressure, $\mathbf{G}$ is the external force, ${\mathbf{F}_s}$ is the surface tension force given by the potential form ${\mathbf{F}_s} = \mu \nabla \phi$. The chemical potential $\mu$ is defined as the variation of the free-energy function of two-phase system expressed by~\cite{Jacqmin}
\begin{eqnarray}
\mu  = 4\beta \phi (\phi  - 1)(\phi-0.5) - k{\nabla ^2}\phi,
\end{eqnarray}
where $k$, $\beta$ are the physical parameters related to the surface tension $\sigma$ and the interface thickness $D$ by $D=\sqrt{8k/\beta}$, $\sigma=\sqrt{2k\beta}/6$.

To solve the Cahn-Hilliard and Navier-Stokes coupled equations, Liang {\it{et al.}}~\cite{Liang5} have presented an improved phase-field-based lattice Boltzmann model by introducing modified equilibrium distribution functions and proper source terms. As a result, the model is capable of recovering the macroscopic equations correctly without any additional assumption and the macroscopic pressure and velocity can be calculated explicitly. The model can also be naturally extended to an efficient three-dimensional version~\cite{Liang6} by using D3Q7 lattice structure for order parameter and D3Q15 lattice structure for flow field, which will be adopted in the current study. The lattice Boltzmann equations with multiple-relaxation-time collision operator can be written as~\cite{Guo}
\begin{equation}
{f_i}\left( {\mathbf{x} + {\mathbf{c}_i}{\delta _t},t + {\delta _t}} \right) - {f_i}(\mathbf{x},t) =  - {({M^{ - 1}}{S_f}M)_{ij}}[{f_j}(\mathbf{x},t) - f_j^{eq}(\mathbf{x},t)] + {\delta _t}{F_i}\left( {\mathbf{x},t} \right),
\end{equation}
\begin{equation}
{{{g}}_i}\left( {\mathbf{x} + {\mathbf{c}_i}{\delta _t},t + {\delta _t}} \right) - {g_i}(\mathbf{x},t) =  - {({\Gamma ^{ - 1}}{S_g}\Gamma )_{ij}}[{g_j}(\mathbf{x},t) - g_j^{eq}(\mathbf{x},t)] + {\delta _t}{G_i}\left( {\mathbf{x},t} \right),
\end{equation}
where ${f_i}$ and $f_i^{eq}$ are the order parameter distribution function and its equilibrium state, ${g_i}$ and $g_i^{eq}$ are the density distribution function and the equilibrium form, $M$ and $\Gamma$ are the collision matrices, ${S_f}$ and ${S_g}$ are the diagonal relaxation matrices, ${F_i}$ and ${G_i}$ are the forcing distribution functions. To recover the correct Cahn-Hilliard equation, the equilibrium distribution function $f_i^{eq}$ is defined by~\cite{Liang5}
\begin{equation}
f_i^{eq}=\left\{
\begin{array}{ll}
\phi+({\omega_i} - 1)\eta\mu,              & \textrm{ $i=0$},    \\
{\omega_i}\eta\mu+{\omega_i}\phi{{{\bf{c}}_i \cdot {\bf{u}}}\over {c_s^2}},  & \textrm{ $i\neq0$} \\
\end{array}
\right.
\end{equation}
and in order to satisfy the divergence-free condition of the velocity, $g_i^{eq}$ is delicately designed as~\cite{Liang5}
\begin{equation}
g_i^{eq}=\left\{
\begin{array}{ll}
{p \over {c_s^2}}({\omega _i} - 1) + \rho{s_i}({\bf{u}}),              & \textrm{ $i=0$},    \\
{p \over {c_s^2}}{\omega _i} + \rho{s_i}({\bf{u}}),                    & \textrm{ $i\neq0$} \\
\end{array}
\right.
\end{equation}
with
\begin{eqnarray}
{s_i}\left(\mathbf{ u} \right) = {\omega _i}\left[ {\frac{{{\mathbf{c}_i} \cdot\mathbf{ u}}}{{c_s^2}} + \frac{{{{\left( {{\mathbf{c}_i} \cdot \mathbf{u}} \right)}^2}}}{{2c_s^4}} - \frac{{\mathbf{u} \cdot \mathbf{u}}}{{2c_s^2}}} \right],
\end{eqnarray}
where ${\omega_i}$ denotes the weighting coefficient, ${c_s}$ represents the speed of sound, $\eta $ is an adjustable parameter for the mobility. The efficient D3Q7 discrete-velocity model is applied to Eq. (4), where the weighting coefficient is given by ${\omega_0}=1/4$, ${\omega_{1-6}}=1/8$, $c_s=c/2$, and the transformation matrix and the discrete velocity are set as~\cite{Liang6}
\begin{eqnarray}
M = \left( {\begin{array}{*{20}{c}}
   1 & 1 & 1 & 1 & 1 & 1 & 1  \\
   0 & 1 & { - 1} & 0 & 0 & 0 & 0  \\
   0 & 0 & 0 & 1 & { - 1} & 0 & 0  \\
   0 & 0 & 0 & 0 & 0 & 1 & { - 1}  \\
   6 & { - 1} & { - 1} & { - 1} & { - 1} & { - 1} & { - 1}  \\
   0 & 2 & 2 & { - 1} & { - 1} & { - 1} & { - 1}  \\
   0 & 0 & 0 & 1 & 1 & { - 1} & { - 1}  \\
\end{array}} \right),
\end{eqnarray}
and
\begin{eqnarray}
{\mathbf{c}_i} = c\left( {\begin{array}{*{20}{c}}
   0 & 1 & { - 1} & 0 & 0 & 0 & 0  \\
   0 & 0 & 0 & 1 & { - 1} & 0 & 0  \\
   0 & 0 & 0 & 0 & 0 & 1 & { - 1}  \\
\end{array}} \right),
\end{eqnarray}
For Eq. (5), we utilize the D3Q15 lattice structure~\cite{dHumieres}, where the weight coefficient is given by ${\omega_0} = 2/9$, ${\omega_{1-6}}=1/9$, ${\omega_{7-14}}=1/72$, ${c_s}=c/\sqrt 3$, the discrete velocity ${\mathbf{c}_i}$ is defined as
\begin{eqnarray}
{\mathbf{c}_i} = c\left( {\begin{array}{*{20}{c}}
   {\rm{0}} & {\rm{1}} & {\rm{0}} & {\rm{0}} & {{\rm{ - 1}}} & {\rm{0}} & {\rm{0}} & {\rm{1}} & {\rm{1}} & {\rm{1}} & {{\rm{ - 1}}} & {{\rm{ - 1}}} & {{\rm{ - 1}}} & {{\rm{ - 1}}} & {\rm{1}}  \\
   {\rm{0}} & {\rm{0}} & {\rm{1}} & {\rm{0}} & {\rm{0}} & {{\rm{ - 1}}} & {\rm{0}} & {\rm{1}} & {\rm{1}} & {{\rm{ - 1}}} & {\rm{1}} & {{\rm{ - 1}}} & {{\rm{ - 1}}} & {\rm{1}} & {{\rm{ - 1}}}  \\
   {\rm{0}} & {\rm{0}} & {\rm{0}} & {\rm{1}} & {\rm{0}} & {\rm{0}} & {{\rm{ - 1}}} & {\rm{1}} & {{\rm{ - 1}}} & {\rm{1}} & {\rm{1}} & {{\rm{ - 1}}} & {\rm{1}} & {{\rm{ - 1}}} & {{\rm{ - 1}}}  \\
\end{array}} \right),
\end{eqnarray}
and the corresponding collision matrix for the D3Q15 model is given as
\begin{eqnarray}
\Gamma {\rm{ = }}\left( {\begin{array}{*{20}{c}}
   {\rm{1}} & {\rm{1}} & {\rm{1}} & {\rm{1}} & {\rm{1}} & {\rm{1}} & {\rm{1}} & {\rm{1}} & {\rm{1}} & {\rm{1}} & {\rm{1}} & {\rm{1}} & {\rm{1}} & {\rm{1}} & {\rm{1}}  \\
   {{\rm{ - 2}}} & {{\rm{ - 1}}} & {{\rm{ - 1}}} & {{\rm{ - 1}}} & {{\rm{ - 1}}} & {{\rm{ - 1}}} & {{\rm{ - 1}}} & {\rm{1}} & {\rm{1}} & {\rm{1}} & {\rm{1}} & {\rm{1}} & {\rm{1}} & {\rm{1}} & {\rm{1}}  \\
   {{\rm{16}}} & {{\rm{ - 4}}} & {{\rm{ - 4}}} & {{\rm{ - 4}}} & {{\rm{ - 4}}} & {{\rm{ - 4}}} & {{\rm{ - 4}}} & {\rm{1}} & {\rm{1}} & {\rm{1}} & {\rm{1}} & {\rm{1}} & {\rm{1}} & {\rm{1}} & {\rm{1}}  \\
   {\rm{0}} & {\rm{1}} & {{\rm{ - 1}}} & {\rm{0}} & {\rm{0}} & {\rm{0}} & {\rm{0}} & {\rm{1}} & {{\rm{ - 1}}} & {\rm{1}} & {{\rm{ - 1}}} & {\rm{1}} & {{\rm{ - 1}}} & {\rm{1}} & {{\rm{ - 1}}}  \\
   {\rm{0}} & {{\rm{ - 4}}} & {\rm{4}} & {\rm{0}} & {\rm{0}} & {\rm{0}} & {\rm{0}} & {\rm{1}} & {{\rm{ - 1}}} & {\rm{1}} & {{\rm{ - 1}}} & {\rm{1}} & {{\rm{ - 1}}} & {\rm{1}} & {{\rm{ - 1}}}  \\
   {\rm{0}} & {\rm{0}} & {\rm{0}} & {\rm{1}} & {{\rm{ - 1}}} & {\rm{0}} & {\rm{0}} & {\rm{1}} & {\rm{1}} & {{\rm{ - 1}}} & {{\rm{ - 1}}} & {\rm{1}} & {\rm{1}} & {{\rm{ - 1}}} & {{\rm{ - 1}}}  \\
   {\rm{0}} & {\rm{0}} & {\rm{0}} & {{\rm{ - 4}}} & {\rm{4}} & {\rm{0}} & {\rm{0}} & {\rm{1}} & {\rm{1}} & {{\rm{ - 1}}} & {{\rm{ - 1}}} & {\rm{1}} & {\rm{1}} & {{\rm{ - 1}}} & {{\rm{ - 1}}}  \\
   {\rm{0}} & {\rm{0}} & {\rm{0}} & {\rm{0}} & {\rm{0}} & {\rm{1}} & {{\rm{ - 1}}} & {\rm{1}} & {\rm{1}} & {\rm{1}} & {\rm{1}} & {{\rm{ - 1}}} & {{\rm{ - 1}}} & {{\rm{ - 1}}} & {{\rm{ - 1}}}  \\
   {\rm{0}} & {\rm{0}} & {\rm{0}} & {\rm{0}} & {\rm{0}} & {{\rm{ - 4}}} & {\rm{4}} & {\rm{1}} & {\rm{1}} & {\rm{1}} & {\rm{1}} & {{\rm{ - 1}}} & {{\rm{ - 1}}} & {{\rm{ - 1}}} & {{\rm{ - 1}}}  \\
   {\rm{0}} & {\rm{2}} & {\rm{2}} & {{\rm{ - 1}}} & {{\rm{ - 1}}} & {{\rm{ - 1}}} & {{\rm{ - 1}}} & {\rm{0}} & {\rm{0}} & {\rm{0}} & {\rm{0}} & {\rm{0}} & {\rm{0}} & {\rm{0}} & {\rm{0}}  \\
   {\rm{0}} & {\rm{0}} & {\rm{0}} & {\rm{1}} & {\rm{1}} & {{\rm{ - 1}}} & {{\rm{ - 1}}} & {\rm{0}} & {\rm{0}} & {\rm{0}} & {\rm{0}} & {\rm{0}} & {\rm{0}} & {\rm{0}} & {\rm{0}}  \\
   {\rm{0}} & {\rm{0}} & {\rm{0}} & {\rm{0}} & {\rm{0}} & {\rm{0}} & {\rm{0}} & {\rm{1}} & {{\rm{ - 1}}} & {{\rm{ - 1}}} & {\rm{1}} & {\rm{1}} & {{\rm{ - 1}}} & {{\rm{ - 1}}} & {\rm{1}}  \\
   {\rm{0}} & {\rm{0}} & {\rm{0}} & {\rm{0}} & {\rm{0}} & {\rm{0}} & {\rm{0}} & {\rm{1}} & {\rm{1}} & {{\rm{ - 1}}} & {{\rm{ - 1}}} & {{\rm{ - 1}}} & {{\rm{ - 1}}} & {\rm{1}} & {\rm{1}}  \\
   {\rm{0}} & {\rm{0}} & {\rm{0}} & {\rm{0}} & {\rm{0}} & {\rm{0}} & {\rm{0}} & {\rm{1}} & {{\rm{ - 1}}} & {\rm{1}} & {{\rm{ - 1}}} & {{\rm{ - 1}}} & {\rm{1}} & {{\rm{ - 1}}} & {\rm{1}}  \\
   {\rm{0}} & {\rm{0}} & {\rm{0}} & {\rm{0}} & {\rm{0}} & {\rm{0}} & {\rm{0}} & {\rm{1}} & {{\rm{ - 1}}} & {{\rm{ - 1}}} & {\rm{1}} & {{\rm{ - 1}}} & {\rm{1}} & {\rm{1}} & {{\rm{ - 1}}}  \\
\end{array}} \right).
\end{eqnarray}
The collision matrix is used to transform the particle distribution function and the equilibrium distribution function into their corresponding moments. With some algebraic manipulations, the equilibrium distribution functions in the moment space can be given by
\begin{eqnarray}
m_f^{eq} = (\phi , \frac{\phi u_x}{c}, \frac{\phi u_y}{c}, \frac{\phi u_z}{c},6\phi-\frac{21\eta\mu}{4},0,0)^T,\hfill\\
m_g^{eq} = {\left( \begin{array}{l}
 0,\frac{{3p + \rho {u^2}}}{{{c^2}}}, - \frac{{45p + 5\rho {u^2}}}{c},\frac{{\rho {u_x}}}{c}, - \frac{{7\rho {u_x}}}{{3c}},\frac{{\rho {u_y}}}{c}, - \frac{{7\rho {u_y}}}{{3c}},\frac{{\rho {u_z}}}{c}, \\
  - \frac{{\rho {u_z}}}{{3c}},\rho \frac{{2u_x^2 - u_y^2 - u_z^2}}{{{c^2}}},\rho \frac{{u_y^2 - u_z^2}}{{{c^2}}},\frac{{\rho {u_x}{u_y}}}{{{c^2}}},\frac{{\rho {u_y}{u_z}}}{{{c^2}}},\frac{{\rho {u_x}{u_z}}}{{{c^2}}},0 \\
 \end{array} \right)^T},
\end{eqnarray}
where ${u_x}$, ${u_y}$ and ${u_z}$ are the ${x-}$, ${y-}$ and ${z-}$ components of macroscopic velocity $\mathbf{u}$. The relaxation matrices ${S_f}$ and ${S_g}$ in Eqs. (4) and (5) are defined by
\begin{eqnarray}
{S_f} = diag\left( {s_0^f,s_1^f, \cdots ,s_6^f} \right),\\
{S_g} = diag\left( {s_0^g,s_1^g,s_2^g,s_3^g \cdots ,s_{14}^g} \right),
\end{eqnarray}
where $0 < s_i^f,~s_i^g< 2$. To derive the Cahn-Hilliard equation correctly, the source term in Eq. (4) is defined by~\cite{Liang5}
\begin{eqnarray}
{F_i} = {\left[ {{M^{ - 1}}\left( {I - \frac{{{S_f}}}{2}} \right)M} \right]_{ij}}\frac{{{\omega _j}{\mathbf{c}_j} \cdot {\partial _t}\phi \mathbf{u}}}{c_s^2},
\end{eqnarray}
and the forcing distribution function ${G_i}$ in Eq. (5) is defined as~\cite{Liang5}
\begin{eqnarray}
{G_i} = {\left[ {{\Gamma ^{ - 1}}({\rm I} - \frac{{{S_g}}}{2})\Gamma } \right]_{ij}} \cdot \frac{{\left( {{\mathbf{c}_j} - \mathbf{u}} \right)}}{{c_s^2}} \cdot \left[ {{s_j}\left(\mathbf{ u} \right)\nabla \left( {\rho c_s^2} \right) + \left( {{\mathbf{F}_s} + {\mathbf{F}_a} + \mathbf{G}} \right)\left( {{s_j}\left( \mathbf{u} \right) + {\omega _j}} \right)} \right],
\end{eqnarray}
where ${\mathbf{F}_a} = \left( {{\rho_h} - {\rho _l}} \right)M{\nabla ^2}\mu \mathbf{u}$ is the additional interfacial force, $\rho_h$, $\rho_l$ represent the densities of the heavy and light fluids. The order parameter $\phi $ can be calculated by
\begin{eqnarray}
\phi  = \sum\limits_i {{f_i}}.
\end{eqnarray}
Once $\phi $ is obtained, the fluid density $\rho $ is determined by
\begin{eqnarray}
\rho  = \phi \left( {{\rho _h} - {\rho _l}} \right) + {\rho _l}.
\end{eqnarray}
The macroscopic pressure and velocity can be obtained by calculating zeroth- or first-order moments of the particle distribution function~\cite{Liang6}
\begin{eqnarray}
\mathbf{u} = \frac{{\sum\limits_i {{\mathbf{c}_i}{{g}_i} + 0.5{\delta _t}\left( {{\mathbf{F}_s} +\mathbf{ G}} \right)} }}{{\rho  - 0.5\left({{\rho_h} - {\rho_l}} \right)M{\nabla ^2}\mu }},\\
p = \frac{{c_s^2}}{{\left( {1 - {\omega _0}} \right)}}\left[ {\sum\limits_{i \ne 0} {{{g}_i} + \frac{{{\delta _t}}}{2}\mathbf{u} \cdot \nabla \rho  + \rho {s_0}\left( \mathbf{u} \right)} } \right].
\end{eqnarray}
The Chapman-Enskog analysis is carried out on the evolution equations (4) and (5), which demonstrates that the mobility $M$ and the kinematic viscosity $\nu$ can be given by~\cite{Liang6}
\begin{eqnarray}
M=\frac{\eta{\delta_x} }{4}\left({\tau_f}-0.5 \right),\\
\nu  = \frac{{\delta_x}}{3}\left( {{\tau _g} - 0.5} \right),
\end{eqnarray}
where ${\tau _f} = 1/s_1^f$, $s_1^f=s_2^f=s_3^f$, ${\tau_g}= 1/{s_9^g}$, $s_9^g =s_{10}^g =s_{11}^g =s_{12}^g=s_{13}^g$. The elements in Eq. (15) are set as $s_1^f=1.25$, $s_4^f=1.2$, $s_0^f=s_5^f=s_6^f=1$, while the relaxation factor $s_9^g$ in Eq. (16) is determined by the given Reynolds number, and the remaining parameters are given as 1. In practice, the time and spatial derivatives need to be evaluated numerically using the following difference schemes,
\begin{eqnarray}
{\partial _t}\chi \left( {\mathbf{x},t} \right) = \frac{{\chi \left( {\mathbf{x},t} \right) - \chi \left( {\mathbf{x},t - {\delta _t}} \right)}}{{{\delta _t}}},
\end{eqnarray}
and
\begin{eqnarray}
\nabla \chi \left( {\mathbf{x},t} \right) = \sum\limits_{i \ne 0} {\frac{{{\omega _i}{\mathbf{c}_i}\chi \left( {\mathbf{x} + {\mathbf{c}_i}{\delta _t},t} \right)}}{{c_s^2{\delta_t}}}},\\
{\nabla ^2}\chi \left( {\mathbf{x},t} \right) = \sum\limits_{i \ne 0} {\frac{{2{\omega _i}\left[ {\chi \left( {x + {\mathbf{c}_i}{\delta_t},t} \right) - \chi \left( {\mathbf{x},t} \right)} \right]}}{{c_s^2\delta _t^2}}},
\end{eqnarray}
where $\chi$ is a related variable.

\section{Numerical Results and Discussions}~\label{sec:Results}
The computational domain consists of a sufficiently long rectangular box divided by an even mesh of ${L_x} \times {L_y} \times {L_z} = W \times W \times 16W$, where $W$ is the width of the box, and a small square-mode perturbation is seeded at the middle plane,
\begin{eqnarray}
h\left( {x,y} \right) = 0.05W\left[ {\cos \left( {kx} \right) + \cos \left( {ky} \right)} \right],
\end{eqnarray}
where $k = 2\pi /W$ is the wave number, and the initial distribution for the order parameter can then be given by
\begin{eqnarray}
\phi \left( {x,y,z} \right) = \tanh 2 {\frac{\left[{z - h\left( {x,y} \right) - 8W}\right]}{D}}.
\end{eqnarray}
Here the dimensionless Atwood ($A_t$) and Reynolds ($Re$) numbers are introduced to describe the evolution of RTI, which are defined as~\cite{Liang3}
\begin{eqnarray}
Re= \frac{{W\sqrt {gW} }}{\nu },~~{A_t} = \frac{{{\rho_h} - {\rho_l}}}{{{\rho_h} + {\rho_l}}},
\end{eqnarray}
where $\nu $ is the kinematic viscosity. To realize the gravitational effect, the following body force $\mathbf{G}$ in the z-direction is applied to the fluids:
\begin{eqnarray}
\mathbf{G} = \left[ {0,0, - \left( {\rho  - \frac{{{\rho_h} + {\rho_l}}}{2}} \right)g} \right].
\end{eqnarray}
Commonly, the characteristic velocity and time are chosen as $\sqrt{gW}$ and $\sqrt{W/g}$, and the following physical quantities have been normalized by these characteristic values. In our simulations, the density of the heavy fluid (${\rho _h}$) is set as 1 and the density for the light fluid (${\rho _l}$) is adjusted according to the setting Atwood number, while the remaining physical parameters are given as: $W=100$, $\sqrt{gW}=0.04$, $\sigma=1\times{10^{-4}}$, $D=4$. The large-scale simulations of 3D long-time RTI become very costly, and to relieve the massive computational cost, all the simulations have been executed using the Graphics Processing Unit parallel technology on the NVIDIA's V100 device. It is shown that the parallel algorithm is capable of deriving a considerable acceleration ratio compared with the CPU machine for simulating the same case.

\subsection{Effect of the Reynolds number}
\begin{figure}
\subfigure[]{\includegraphics[width=6cm]{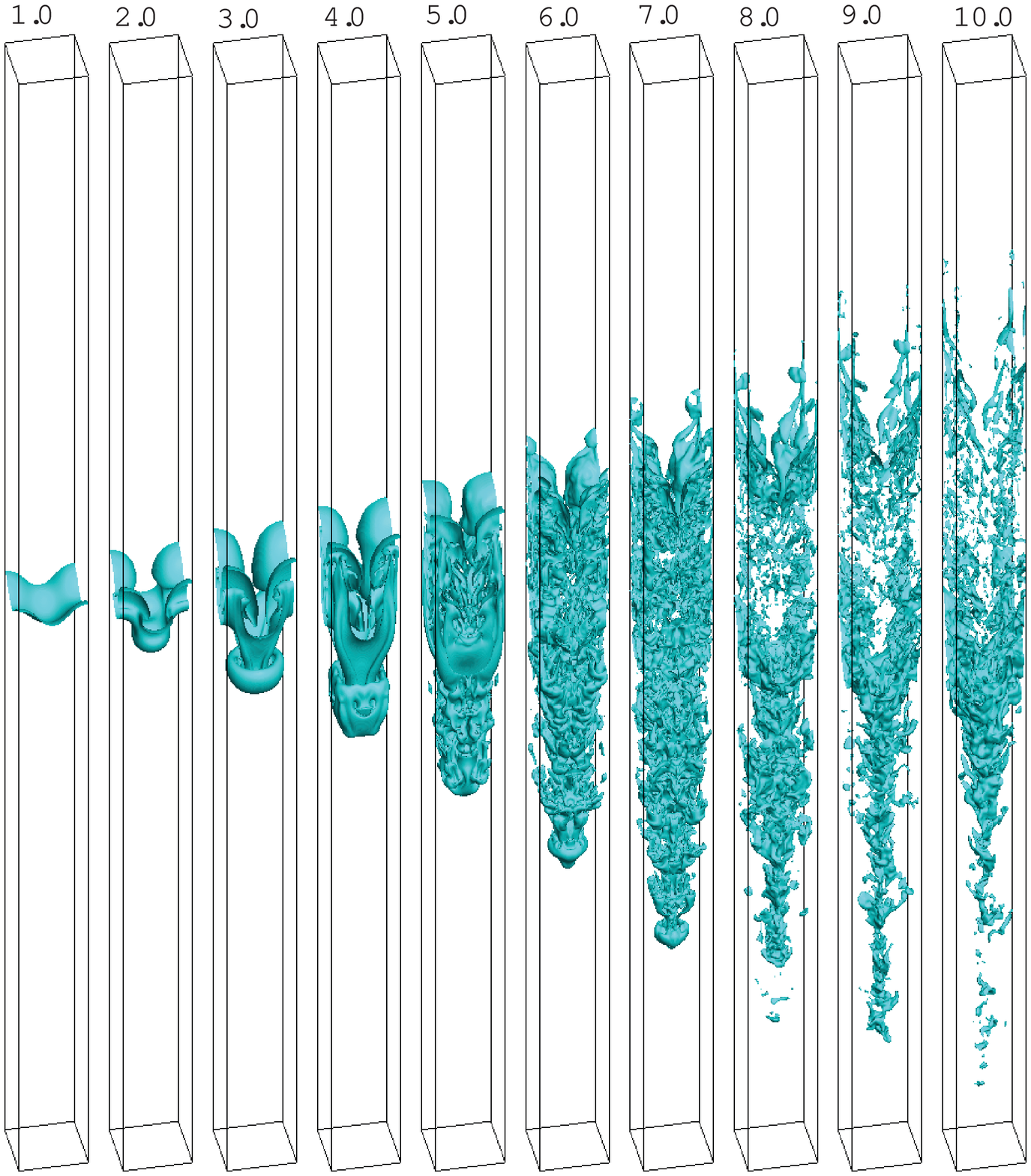}}
\subfigure[]{\includegraphics[width=6cm]{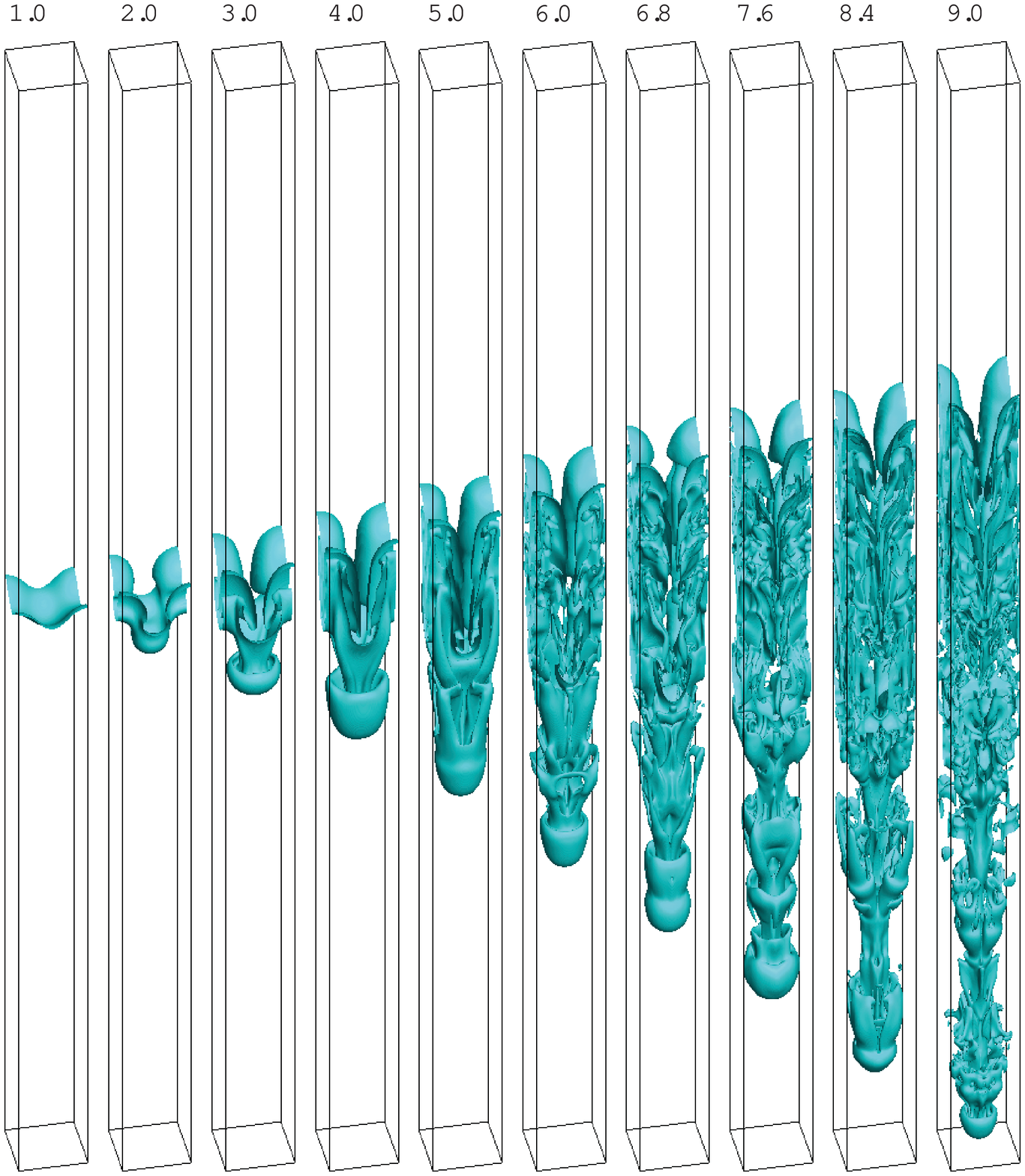}}\\
\subfigure[]{\includegraphics[width=6cm]{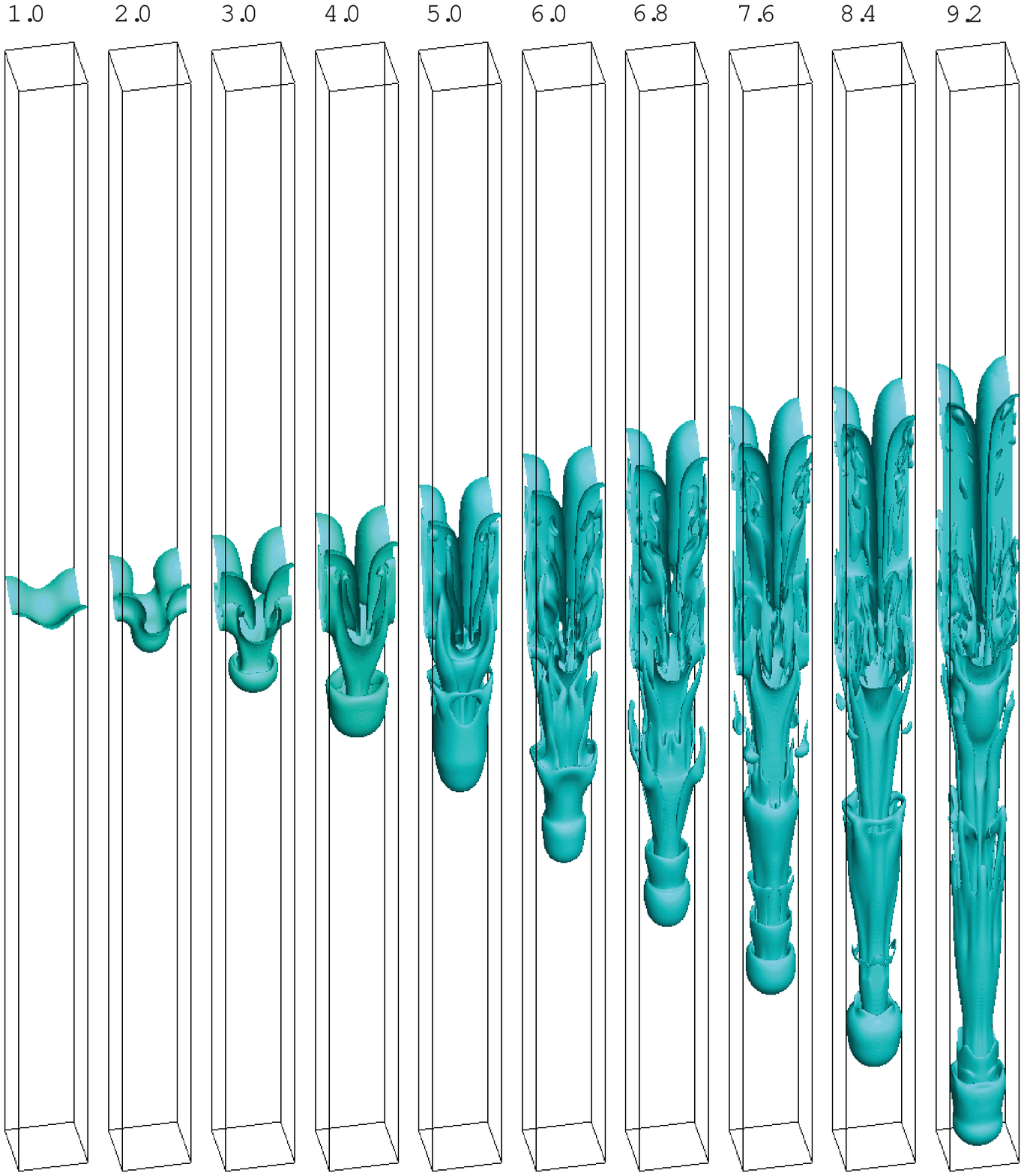}}
\subfigure[]{\includegraphics[width=6cm]{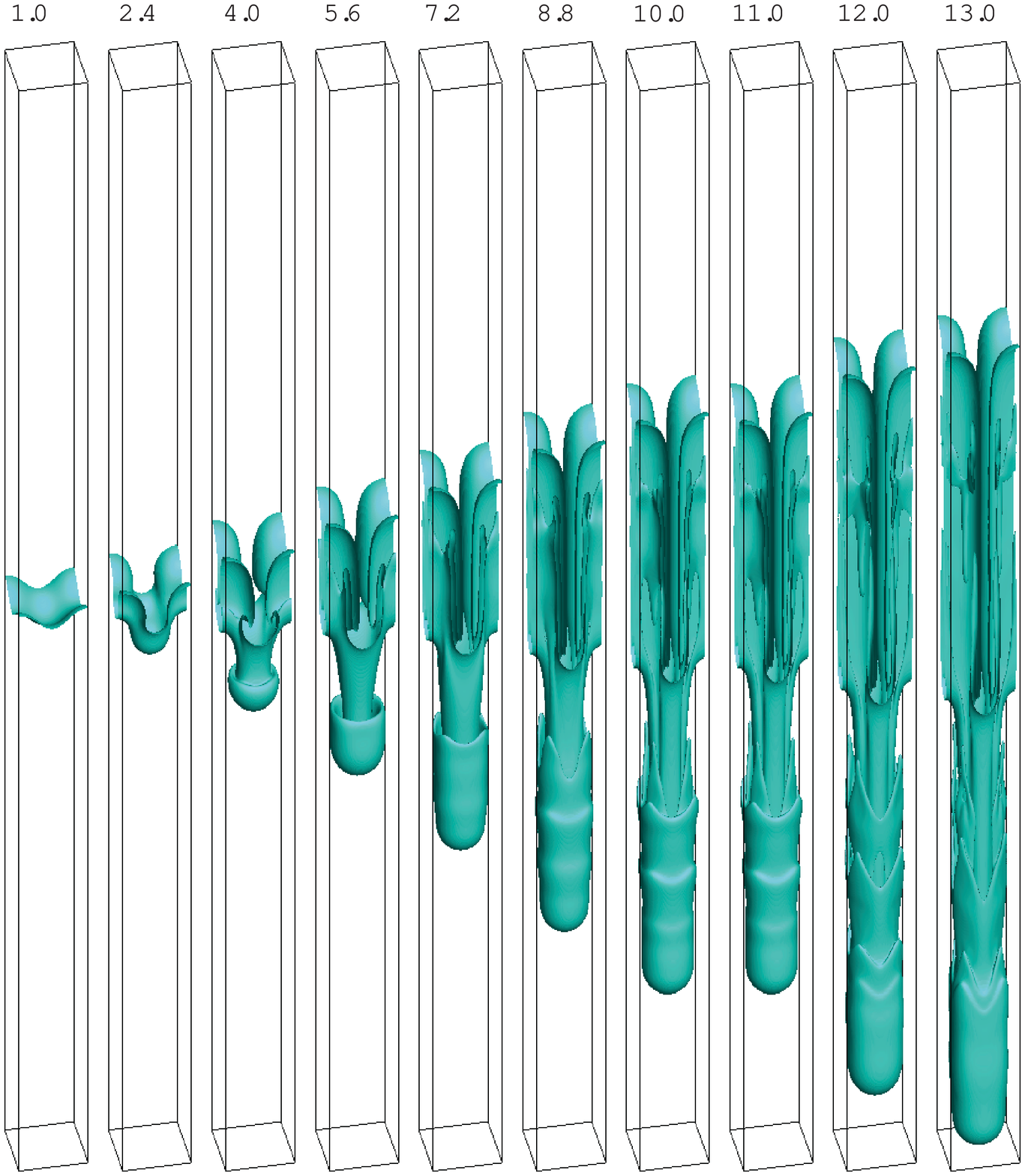}}
\caption{Time evolution of phase interface in 3D single-mode RTI with various values of $Re$, $A_t=0.5$: (a) $Re=5000$, (b) $Re=1000$, (c) $Re=500$, (d) $Re=100$.}
\end{figure}

\begin{figure}
\subfigure[]{\includegraphics[width=6cm]{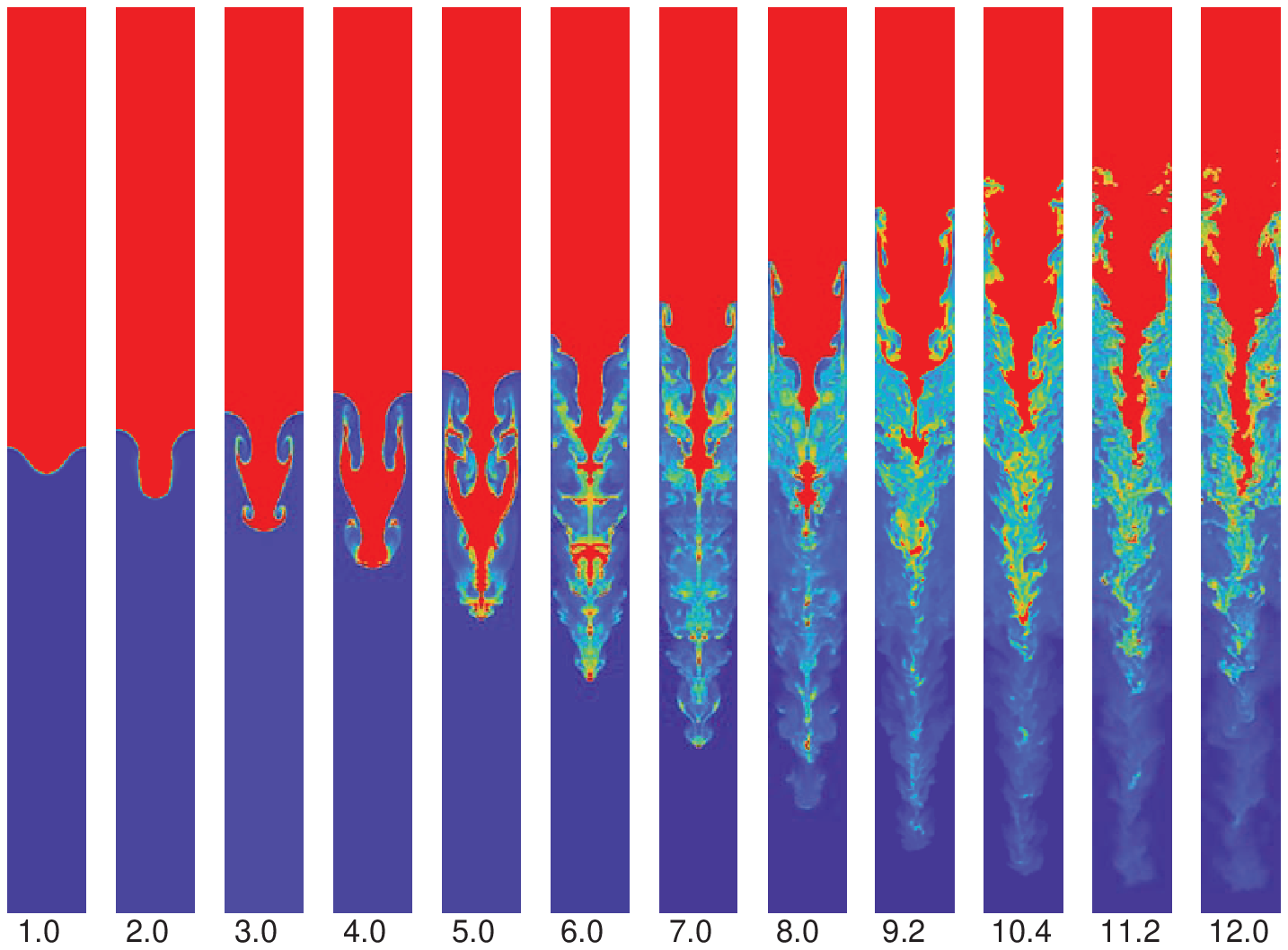}}
\subfigure[]{\includegraphics[width=6cm]{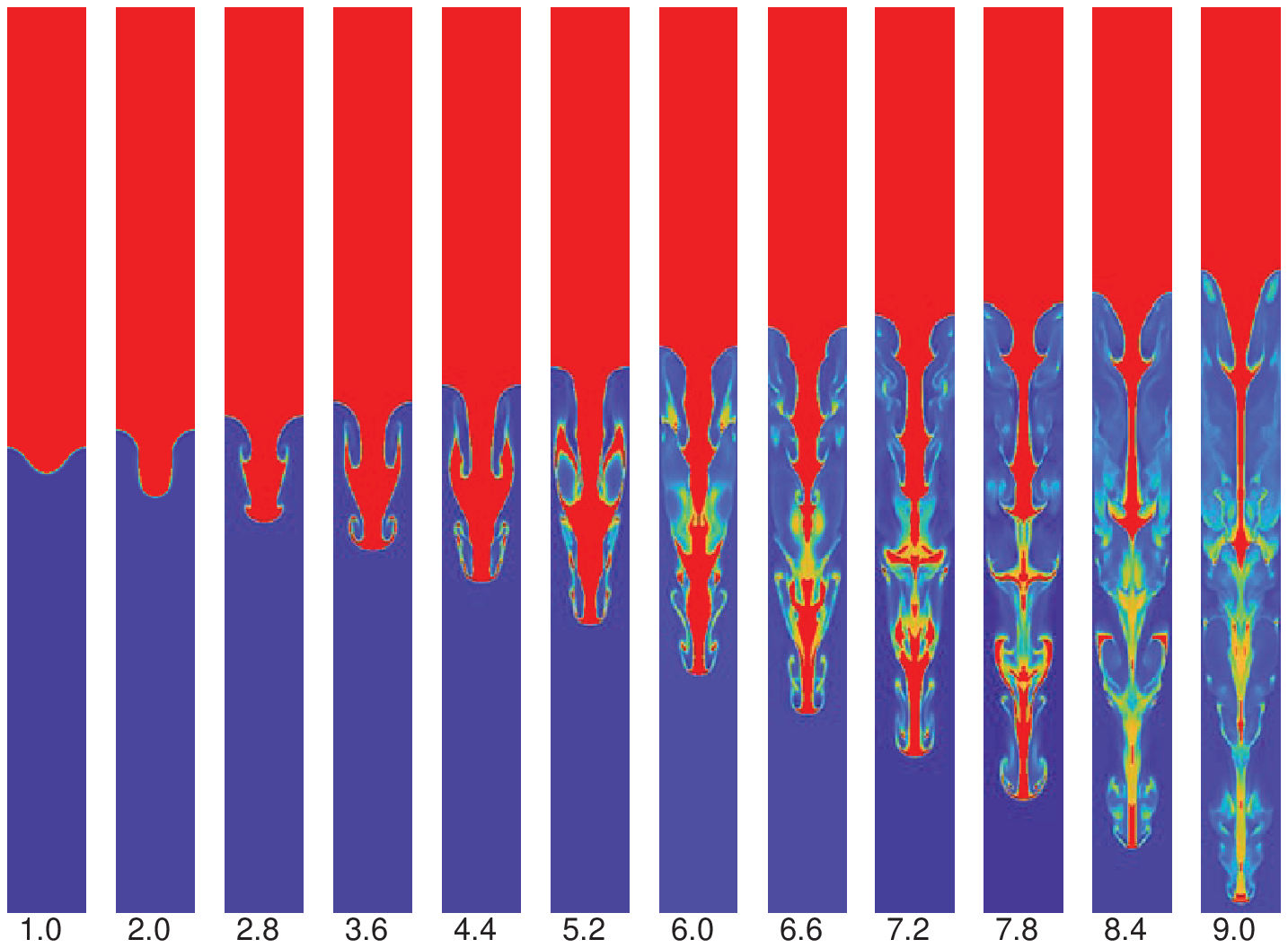}}\\
\subfigure[] {\includegraphics[width=6cm]{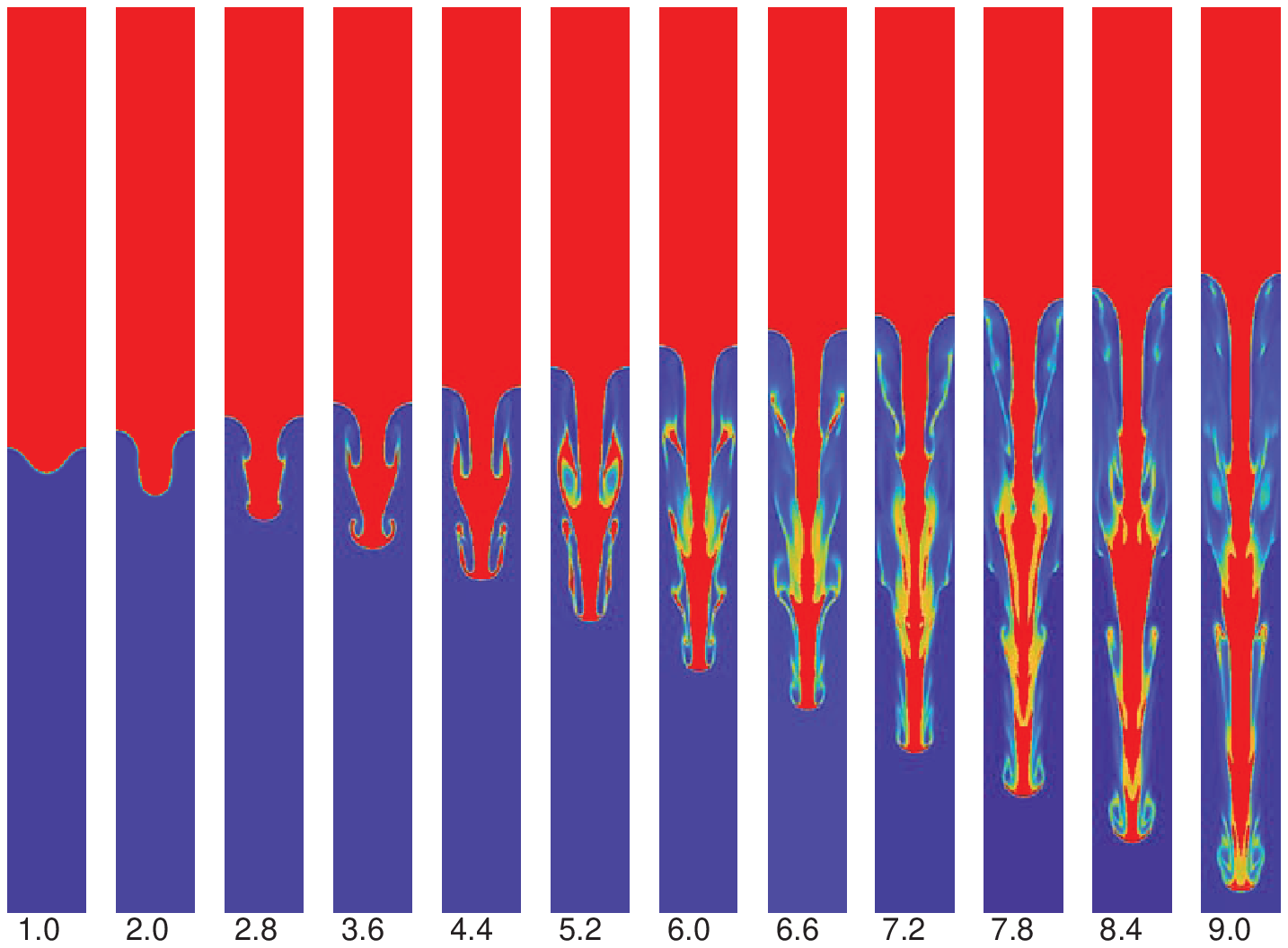}}
\subfigure[] {\includegraphics[width=6cm]{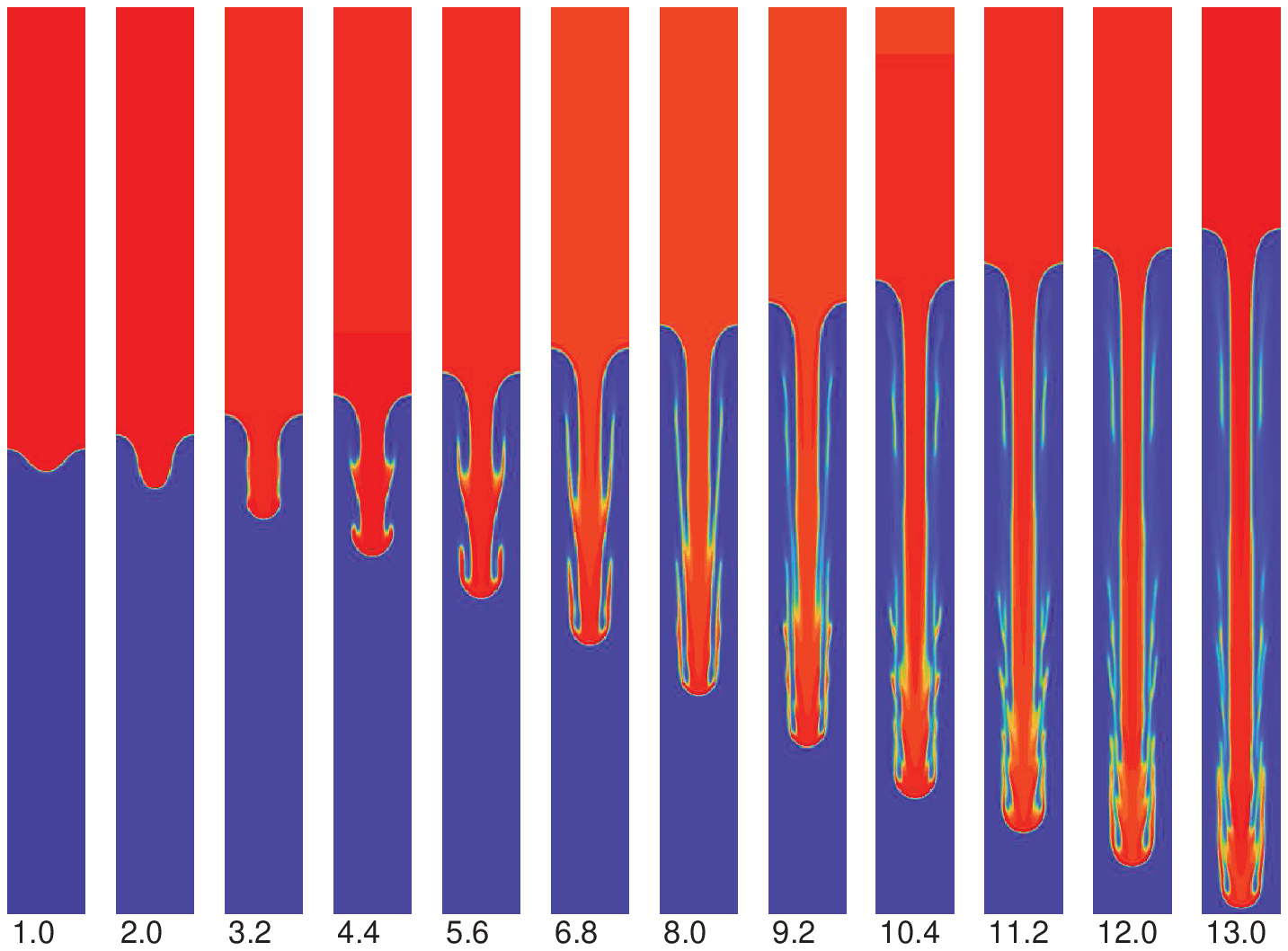}}
\caption{Time evolution of the density image at the diagonal vertical plane, $A_t=0.5$: (a) $Re=5000$, (b) $Re=1000$, (c) $Re=500$, (d) $Re=100$.}
\end{figure}

The influence of the Reynolds number on the evolution of single-mode RTI is first studied at a medium Atwood number of 0.5. Figure 1 shows the time evolution of interfacial pattern in 3D single-mode RTI with four representative Reynolds numbers. It can be observed from Fig. 1 that the instability under each Reynolds number displays similar characteristic at the initial stage: the heavy fluid descends to form a spike and the light fluid rises to form bubble. Then the spike rolls up at its tail and a classic mushroom-like structure can be observed owing to the action of the nonlinear Kelvin-Helmholtz instability. The mushroom structure continues to increase in size and then displays diverse behaviours at different Reynolds numbers. For a high Reynolds number of 5000, the interface rolls up at multiple positions and we can see the produced new spikes and bubbles which is equivalent to the occurrence of secondary RTI. As time advances, the strengthes of Kelvin-Helmholtz vortices increase continuously leading to the severe interfacial deformation and some of the interfaces even undergo the chaotic breakups. Finally, some dissociative drops can be clearly observed in the system. In addition, we can inspect that the interface structure at the late-time evolutional process breaks the symmetry with respect to the middle axis, which can also be shown in the corresponding 2D plane of Fig. 2(a). As the Reynolds number is reduced gradually, the complexity of the interface at the late time is weakened accordingly and the phase interface in the whole process becomes more and more smooth. In particular, the heavy fluid at a low Reynolds number of 100 falls down continuously in the manner of the spike without the appearance of the breakup phenomenon. Besides, it can also be observed that the symmetry of interfacial pattern can be always maintained for the Reynolds number lower than 1000. To show the interfacial dynamics more clearly, we also plotted in Fig. 2 the time evolution of the density image at the diagonal vertical section with the above Reynolds numbers. We can observe a unique phenomenon of 3D single-mode RTI compared to the 2D example that is the formation of two pairs of counter-rotating vortices, and the size of vortex increases with the Reynolds number. For a high Reynolds number, the new or secondary roll-up behaviours then occur at the multiple layers and numerous vortices at different scales are produced. At the late time of the evolution, the interaction of fluids in the mixing zone becomes more and more intense, which leads to the rupture of many vortices and eventually inducing the fully turbulent mixing of fluids. Also, the symmetry of interfacial pattern is obviously destroyed. The multiple-layer roll-ups can also be observed at the moderate Reynolds number, while it appear at later time and the vortex sizes are substantially smaller. When the Reynolds number is very small, the evolution of the phase interface presents a laminar flow state. This is because the influence of the viscous force on the flow field is greater than the inertia, and the disturbance of the flow field will be attenuated by the viscous force enforcing a relatively stable shear layer.

\begin{figure}
\subfigure[]{\includegraphics[width=3.0in,height=2.5in]{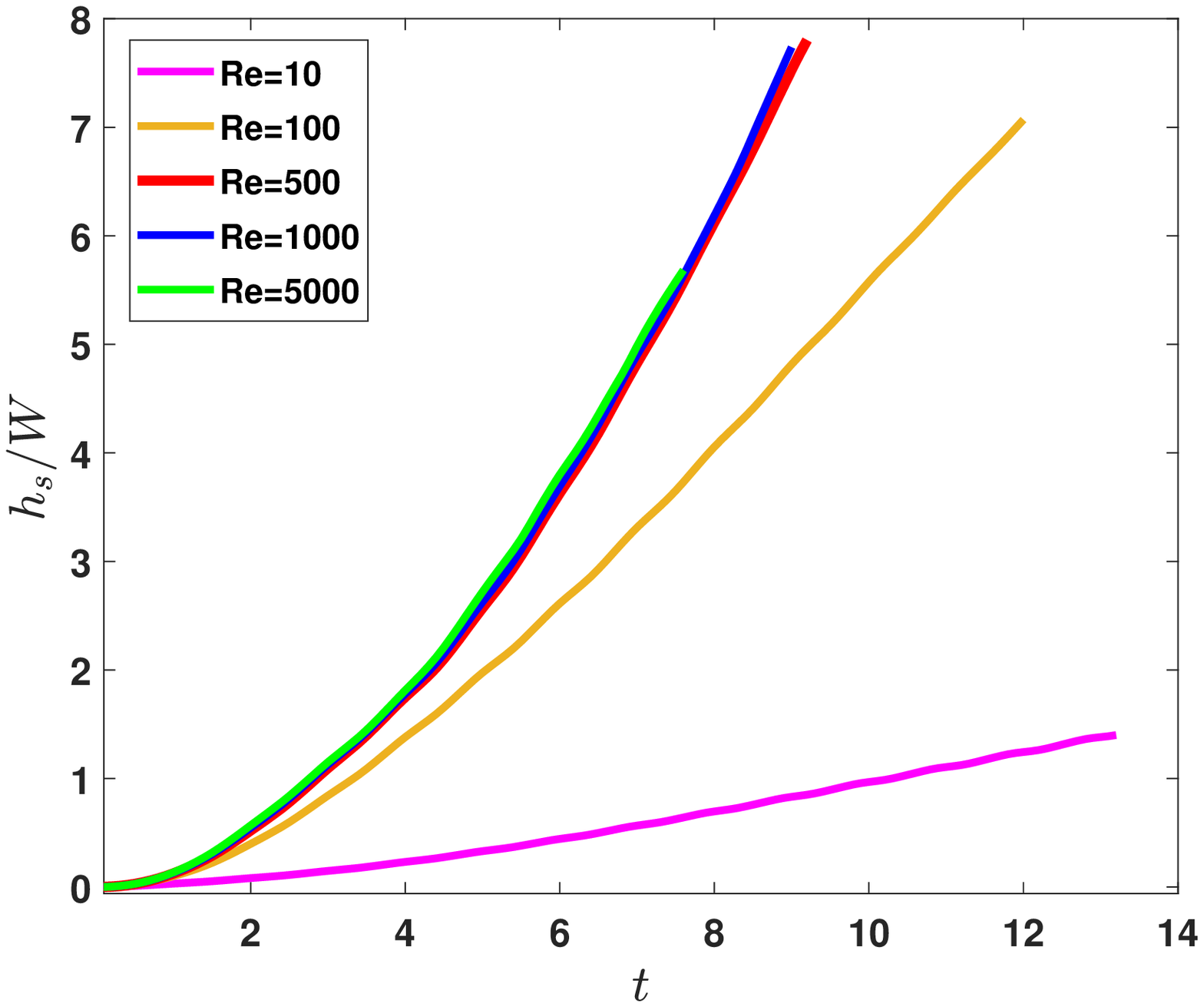}}
\subfigure[]{\includegraphics[width=3.0in,height=2.5in]{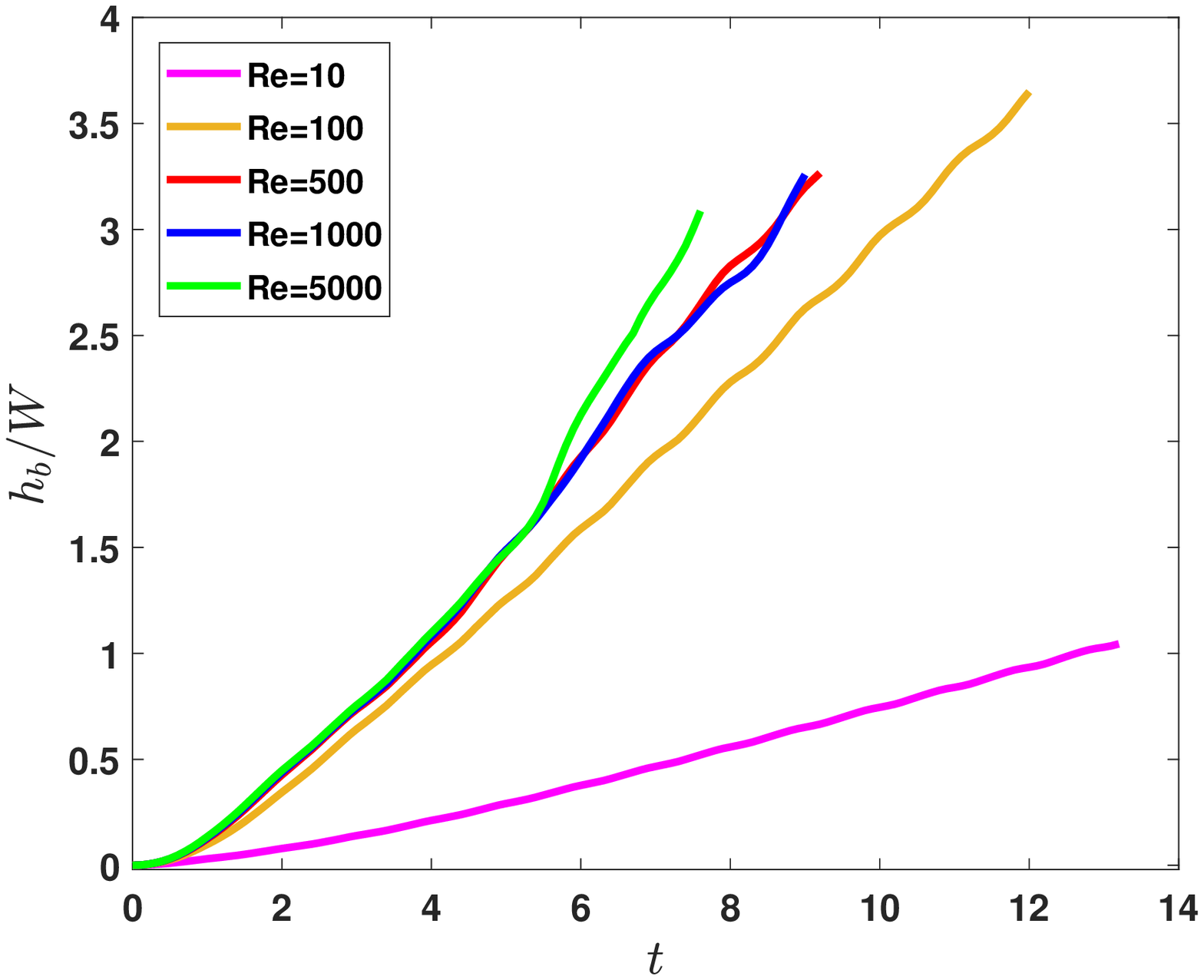}}
\caption{Effect of Reynolds number on the normalized (a) spike and (b) bubble amplitudes of 3D single-mode RTI.}
\end{figure}

\begin{figure}
\subfigure[]{\includegraphics[width=3.0in,height=2.5in]{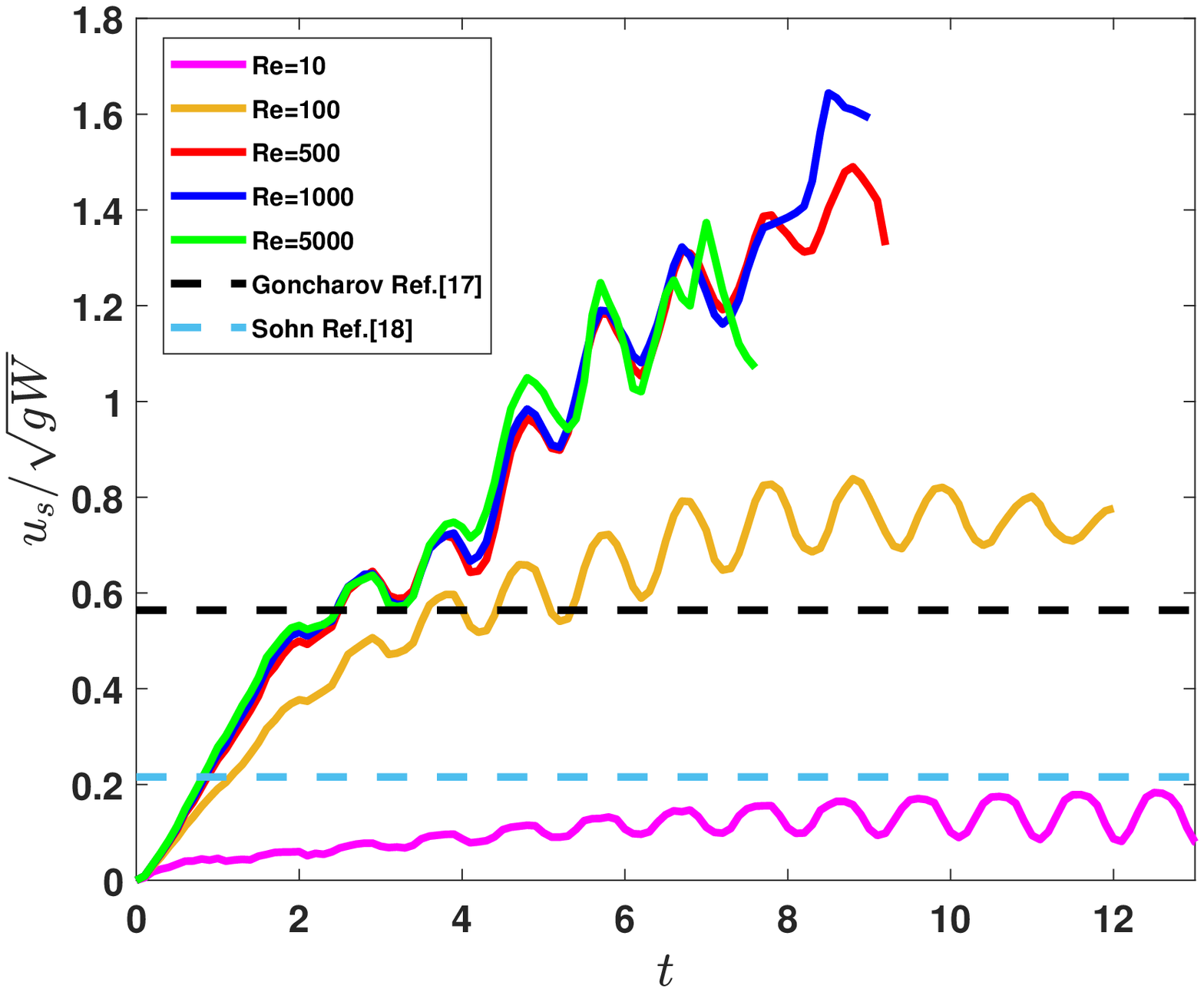}}
\subfigure[]{\includegraphics[width=3.0in,height=2.5in]{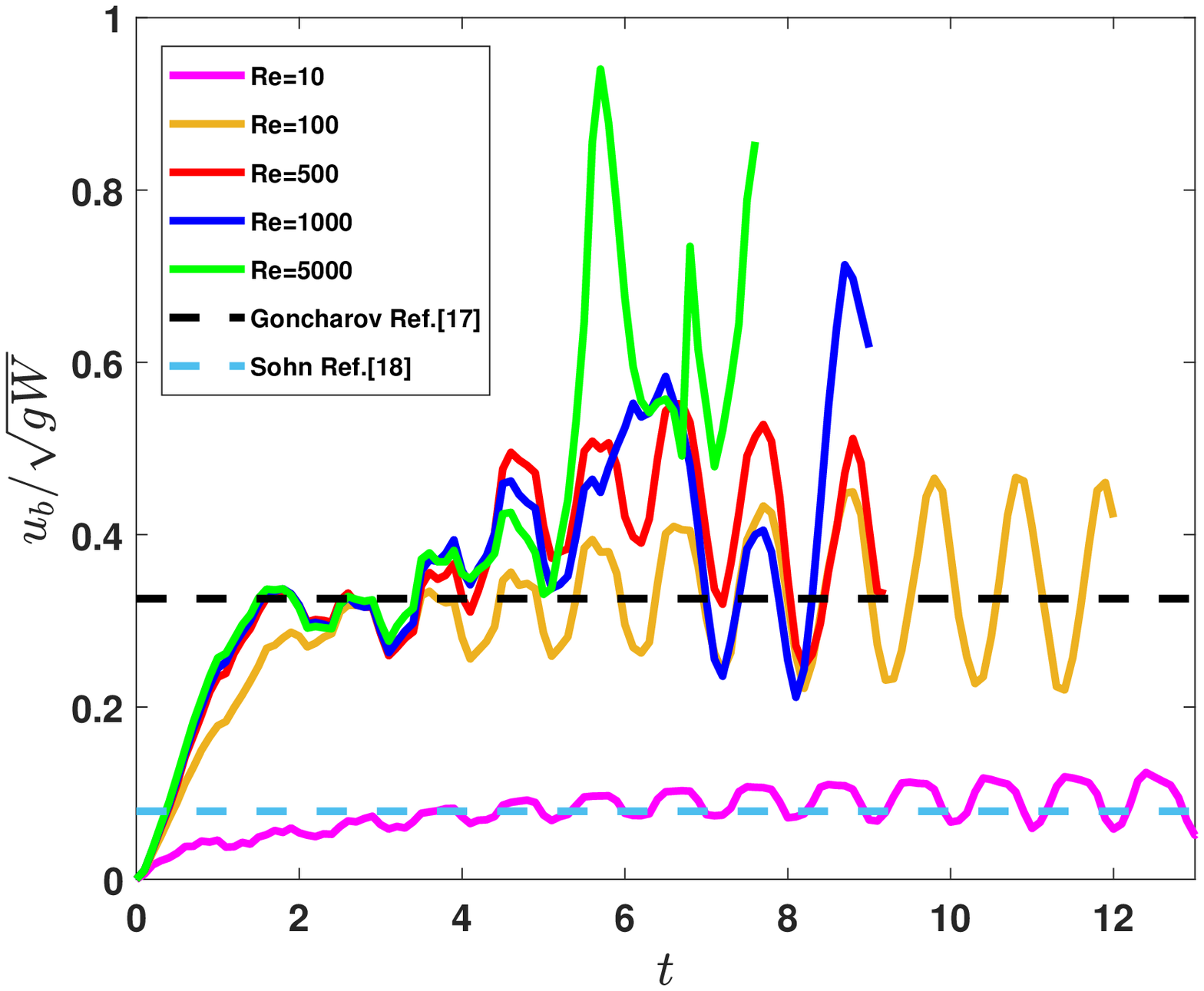}}
\caption{Effect of Reynolds number on the (a) normalized spike velocity and (b) normalized bubble velocity in 3D single-mode RTI. The black dotted line denotes the analytical solution of the potential flow model proposed by Goncharov~\cite{Goncharov}, while the blue dashed line marks the analytical solution of the model proposed by Sohn~\cite{Sohn}.}
\end{figure}

We also computed the normalized amplitudes of the spike and bubble and presented in Fig. 3 their evolutional curves with extensive Reynolds numbers. From Fig. 3, it can be observed that the spikes and bubble amplitudes increase with time, and the amplitude slopes for both the spike and bubble also increase with the Reynolds number. However, the increase ranges in the amplitude slopes slow down with the Reynolds number, and it is evident in Fig. 3 that the curves of bubble and spike amplitudes have the trends of coincidence when the Reynolds number is large enough. We theoretically analyzed the relationship between the Reynolds number and the disturbance amplitude. The spike dynamics of incompressible RTI can be determined by the relative importance of the buoyancy and dissipation forces of per unit mass~\cite{Abarzhi}
\begin{eqnarray}
\frac{{{{d}}{h_s}}}{{dt}} = {u_s},~~\frac{{d{u_s}}}{{dt}} = {A_t}g + {F},
\end{eqnarray}
where $h_s$ is the spike amplitude, $u_s$ is the spike velocity, the dissipation force is the rate of momentum loss in the direction of gravity and ${F} =-\varepsilon/\nu$, $\varepsilon $ is the energy dissipation rate with the neglect of the viscous time scale can be expressed by ${\varepsilon} = {C}{\nu^3}/W$~\cite{Sreenivasan}, then the viscous dissipation force can be give as ${F} = -{C}{v^2}/W$, ${C}$ is a positive constant. From the above analysis, it indicates that the viscous dissipation force as a resistance decreases with the increase of the Reynolds number such that it could promote the growth of spike front. As the Reynolds number is increased to be sufficiently large, the viscous dissipation force compared to the buoyancy force is negligibly small, thus the instability growth would be little dependence of the Reynolds number. The comparison between the theoretical analysis and our numerical simulation shows a good agreement.

Figure 4 shows the normalized velocities of the spike and bubble versus time under different Reynolds numbers. Based on the velocity curves, we can identify the development of 3D single-mode RTI at a high Reynolds number into four different stages: linear growth stage, saturated velocity growth stage, reacceleration stage and turbulent mixing stage. After the initial linear stage, the spike and bubble grows with approximately constant velocities as shown in Fig. 4, although the duration of this stage for the spike is very shorter than that of the bubble. Goncharov~\cite{Goncharov} proposed an analytical potential flow model for predicting the constant spike and bubble velocities,
\begin{equation}
u_s=\sqrt{\frac{2A_tg}{(1-A_t)k}},~~u_b=\sqrt{\frac{2A_tg}{(1+A_t)k}}.
\end{equation}
We also compared the simulation results with this theoretical solutions of the potential flow model in Fig. 4, and good agreements between them can be achieved. In the following, the strengthes of the nonlinear vortices increase gradually, which drives the velocities of the spike and bubble exceeding the asymptotic values of the potential flow theory~\cite{Goncharov}. This implies that the evolution of the stability has entered into the reacceleration stage. The reacceleration stage does not last forever and the flow would be converted into the chaotic state at the late time. In the turbulent mixing stage, the evolution speeds of the spike and bubble become unstable with the increasing complexity of the vortex structures, and exhibit the deceleration and acceleration cycle. As the Reynolds number decreases, the flow instability is reduced and the late-time stages including the chaotic development stage and the reacceleration stage cannot be successively reached. For example, the spike velocity does not reach the turbulent mixing stage at a Reynolds number of 100, and the evolutions of the spike and bubble end to the saturated velocity growth stage when Reynolds number is 10. In addition, we also noted that the spike and bubble quasisteady velocities are smaller than the potential flow theory of
Goncharov\cite{Goncharov}, due to the ignored viscous effect in his analysis. Later, Sohn~\cite{Sohn} incorporated the effects of the fluid viscosity and surface tension into the saturated velocities of the bubble and spike as
\begin{eqnarray}
{{{u}}_{{{b}},~s}} = \sqrt {\frac{{2A_tg}}{{\left( {1 \pm A_t} \right)k}} - \frac{{3k\sigma }}{{16{\rho _h}}} + {k^2}{v^2}} - kv,
\end{eqnarray}
In Fig. 4, we also plotted the analytical solutions of the modified potential flow model at a low Reynolds number of 10, and good agreements between the numerical results and the analytical solutions are achieved in general.

\begin{figure}
\centering
\subfigure[]{\includegraphics[width=3.0in,height=2.5in]{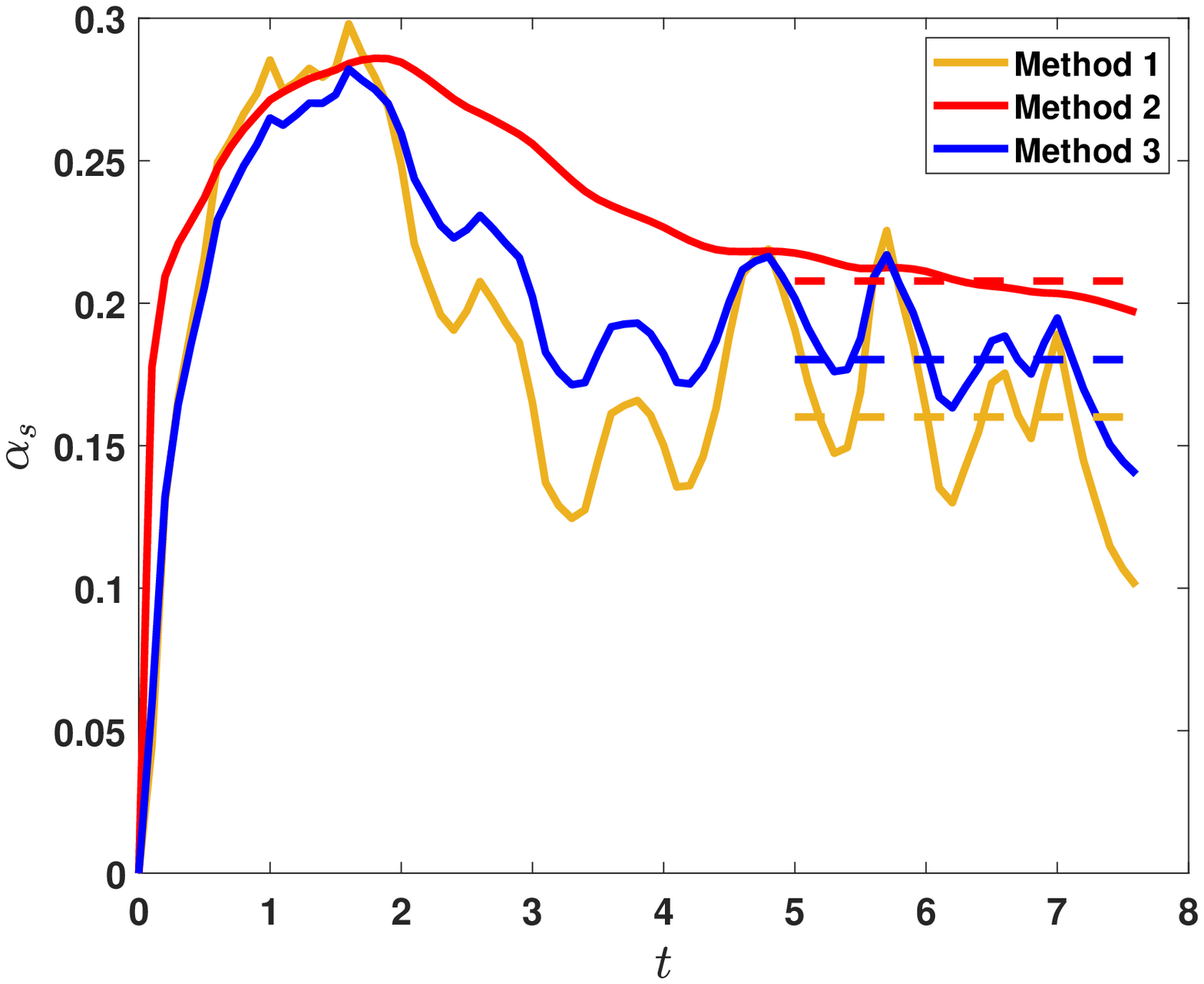}}
\subfigure[]{\includegraphics[width=3.0in,height=2.5in]{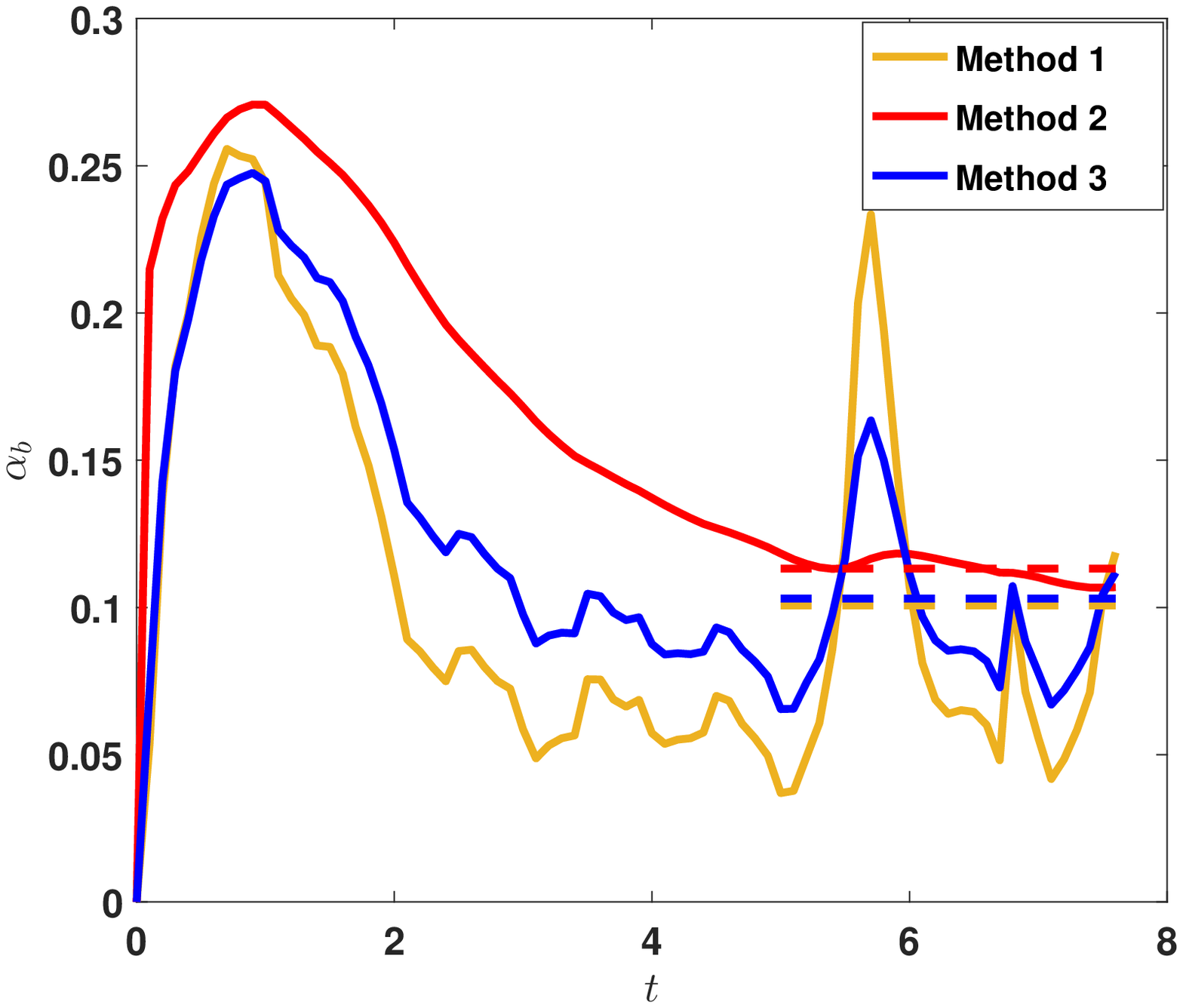}}
\caption{Time variations of the (a) spike and (b) bubble normalized accelerations computed by Methods 1, 2 and 3. The dotted lines represent the averages of the spike and bubble accelerations at turbulent mixing stage.}
\end{figure}

\begin{figure}
\subfigure[]{\includegraphics[width=3.0in,height=2.5in]{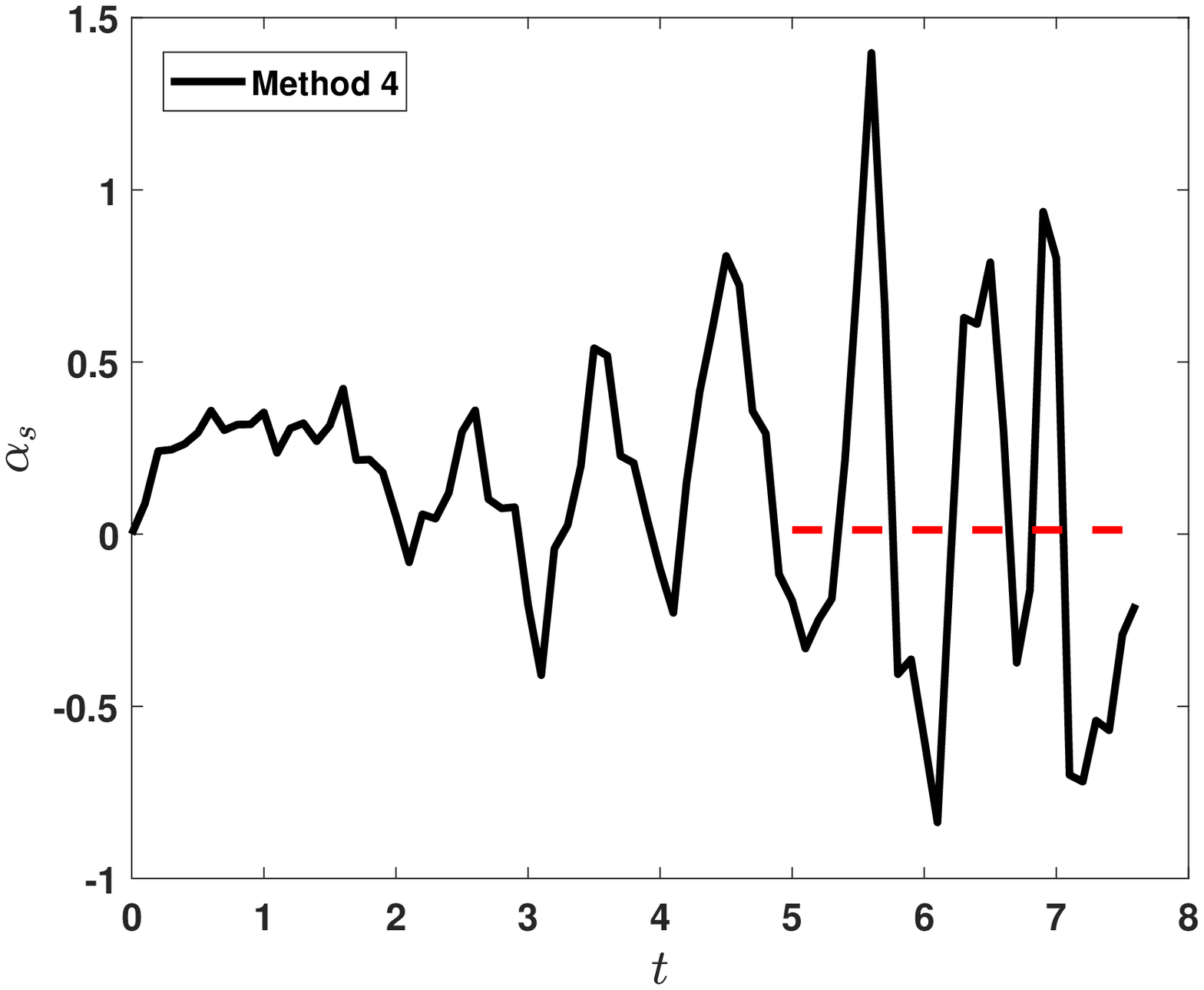}}
\subfigure[]{\includegraphics[width=3.0in,height=2.5in]{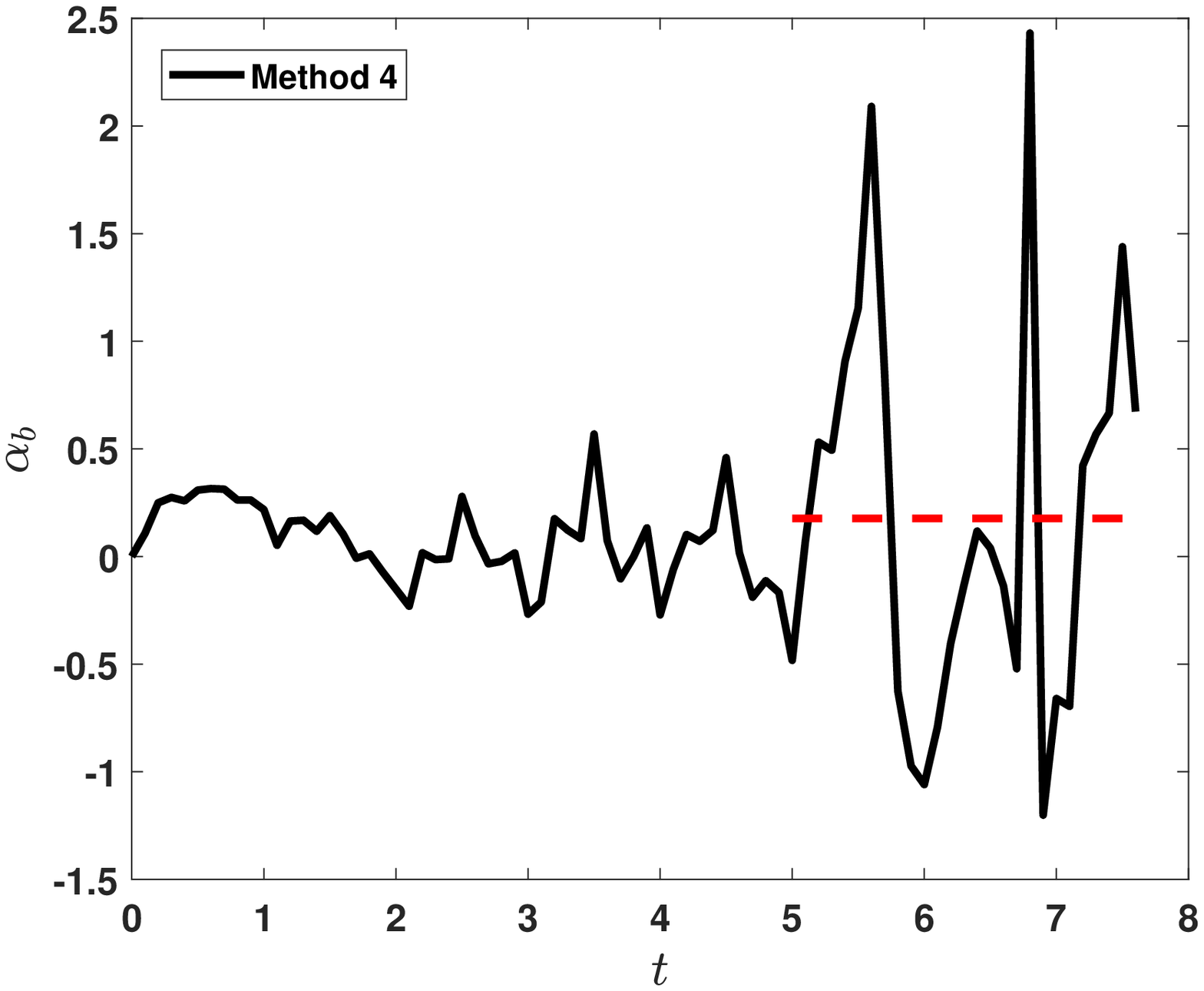}}\\
\subfigure[]{\includegraphics[width=3.0in,height=2.5in]{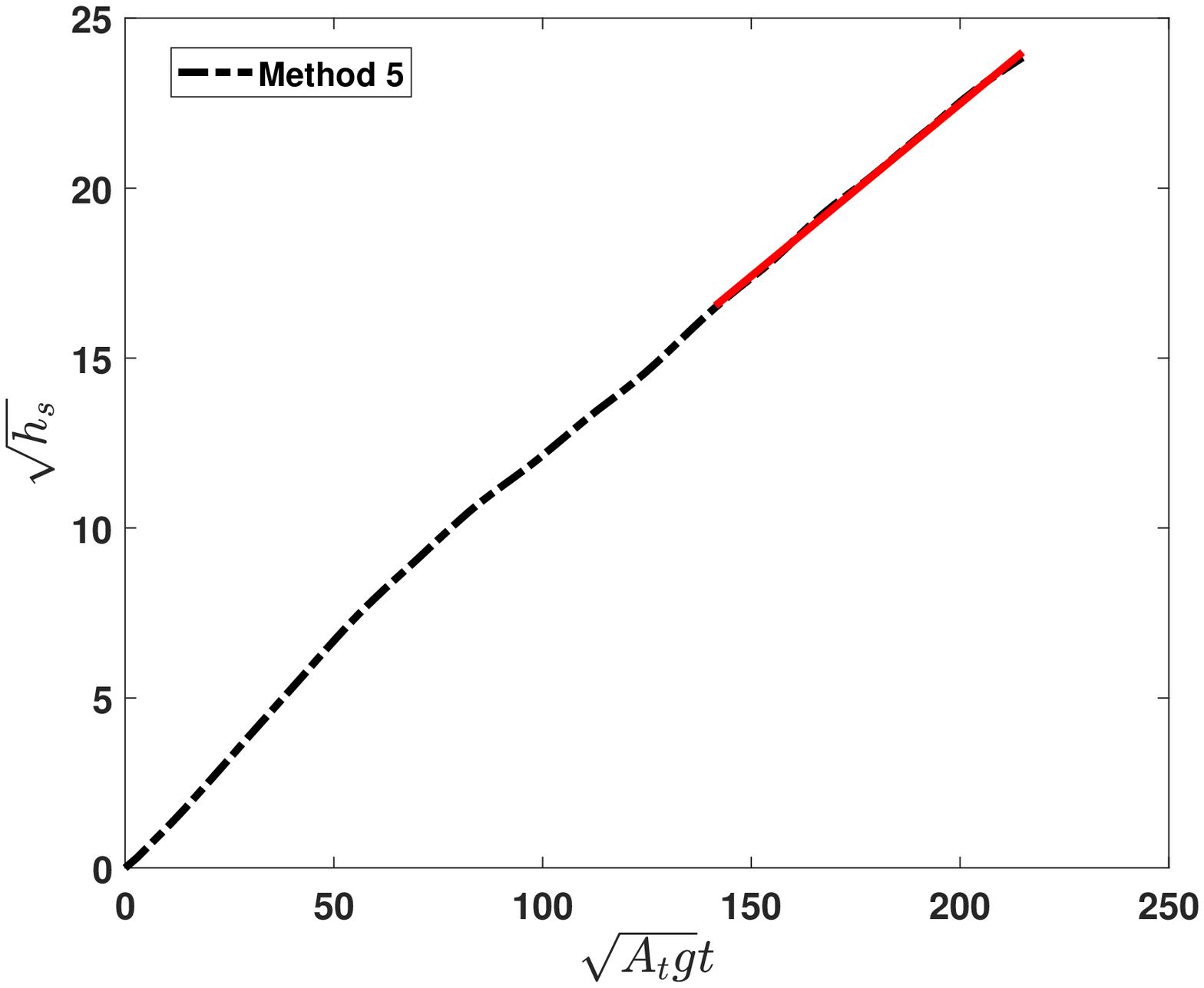}}
\subfigure[]{\includegraphics[width=3.0in,height=2.5in]{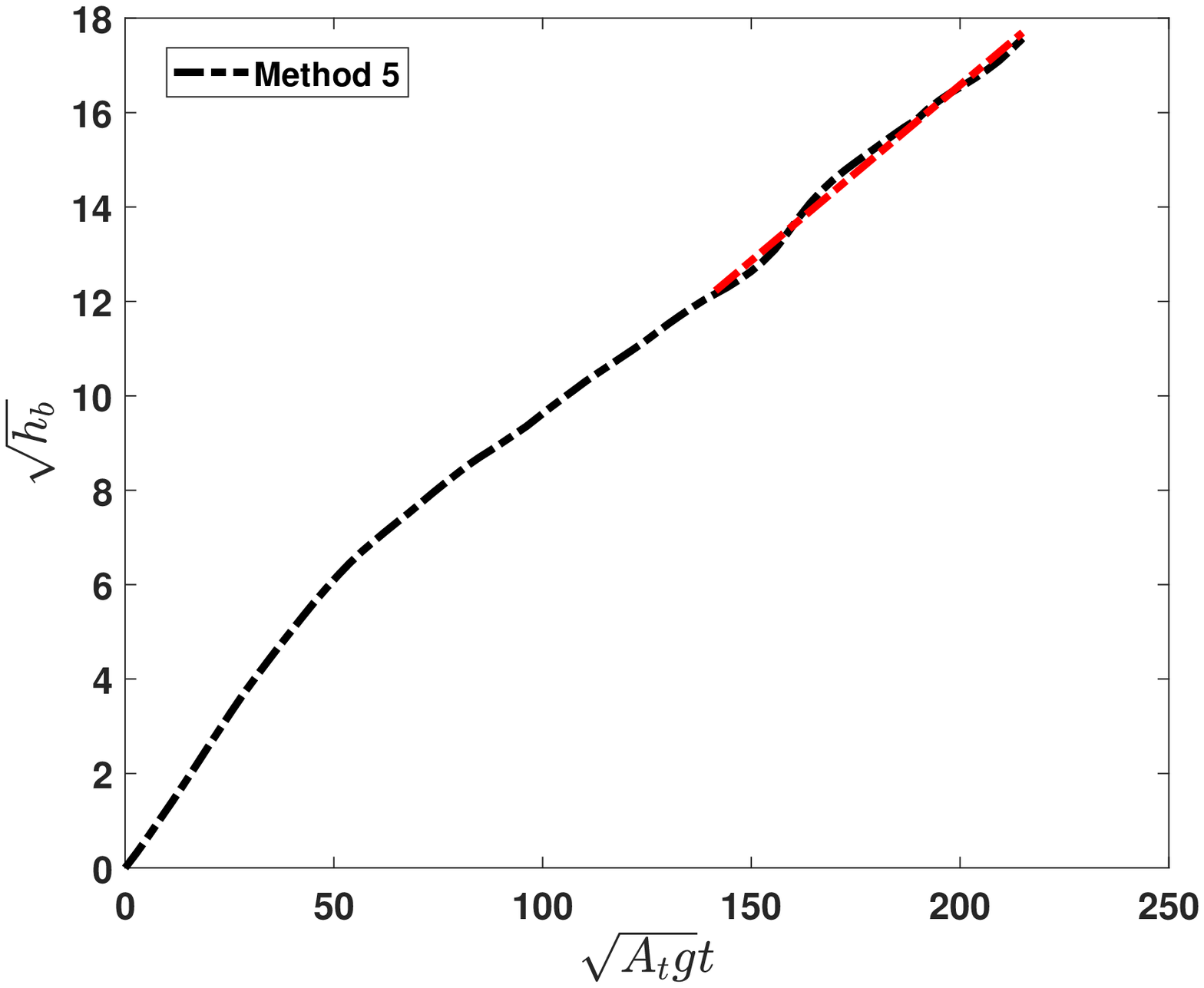}}
\caption{Time variations of the (a) spike and (b) bubble normalized accelerations computed by Method 4 and the dotted lines are the averaging values at the late-time stage. The dependence of (c) $\sqrt {{{{h}}_s}} $ and (d) $\sqrt{{{{h}}_b}} $ on the variable $\sqrt{{A_t}g} t$ by Method 5 and the solid lines represent their late-time linear fitting curves.}
\end{figure}

According to the latest researches~\cite{Wei, Liang1}, it is generally accepted that the amplitude of single-mode RTI at a high Reynolds number has a quadratic growth law in the turbulent mixing stage, i.e., $h_{s, b}=\alpha_{s,b}{A_t}gt^2$, where $\alpha_{s,b}$ are the spike and bubble growth rates. Several statistical methods~\cite{Ristorcelli, Cook, Cabot, Clark, Wei, Jacobs} have been originally proposed for computing the growth rate of Rayleigh-Taylor turbulence and was recently adopted to calculate the spike and bubble growth rates of 2D single-mode RTI. However, these methods have not been applied to the 3D situation, thus a detailed comparison among them was still conducted here. Based on the fact $\frac{d h_{s,b}}{dt}=2\alpha_{s,b}{A_t}gt$, the first method for computing the spike and bubble growth rates is presented as~\cite{Ristorcelli, Cook}
\begin{eqnarray}
\alpha_{s,b} = \frac{\dot h_{s,b}^2}{4{A_t}g{h_{s,b}}},
\end{eqnarray}
where ${h_{s,b}}$ with the subscripts $s$ and $b$ represent the time derivatives of the spike and bubble amplitudes. The second method to determine the late-time growth rates can be directly derived as~\cite{Cabot}
\begin{eqnarray}
{\alpha _{s,b}} = \frac{{{h_{s,b}}}}{{{A_t}g{t^2}}}.
\end{eqnarray}
Clark~\cite{Clark} used a lattice Boltzmann multiphase algorithm to investigate the immiscible Rayleigh-Taylor mixing and proposed a third method for determining the growth rates of the mixed layer,
\begin{eqnarray}
{\alpha _{s,b}} = \frac{{\partial {h_{s,b}}}}{{\partial Z}} = \frac{{\partial {h_{s,b}}}}{{\partial z}}\frac{{\partial z}}{{\partial Z}} = \frac{1}{{2A_t}}\frac{{\partial {h_{s,b}}}}{{\partial z}},
\end{eqnarray}
where $Z = 2{A_t}z$, $z =g{t^2}/2$. In addition, Wei~\cite{Wei} proposed the fourth method for measuring the growth rates in a straight manner,
\begin{eqnarray}
{\alpha _{s,b}} = \frac{{{{\ddot h}_{s,b}}}}{{2{A_t}g}}.
\end{eqnarray}
Alternatively, Olson and Jacobs~\cite{Jacobs} took the square root of the equation ${h_{s,b}} = {\alpha _{s,b}}Ag{t^2}$, resulting in the fifth measurement method
\begin{eqnarray}
h_{s,b}^{1/2} = {\left( {\alpha_{s,b} {A_t} g} \right)^{1/2}}t,
\end{eqnarray}
where the growth rates ${\alpha_{s,b}}$ can be obtained by plotting the linear fitting curves between ${h_{s,b}^{1/2}}$ and ${\left({A_tg}\right)^{1/2}}t$. Figure 5 depicts the time evolutions of the acceleration coefficients of the spike and bubble predicted by Methods 1, 2 and 3. It can be found that the acceleration curves calculated by Method 2 is relatively smooth with slight reductions by the end of the simulation, and the predicted spike and bubble accelerations using Methods 1 and 3 have some fluctuations in the turbulent mixing stage. We presented the late-time averages of the spike and bubble acceleration coefficients in Table I, which indicates that the spike growth rates by Method 1, 2, and 3 are 0.1601, 0.2060, and 0.1790, respectively, while the corresponding predictions for the bubble are 0.1008, 0.1127 and 0.1030. Figures 6(a) and 6(b) depict the time variations of normalized accelerations of the spike and bubble computed by Method 4 and we can find that their late-time growth rates fluctuate around the mean values of 0.0489 and 0.1476, respectively. In Figs. 6(c) and 6(d), the relations between $\sqrt {{h_{s,b}}} $ and $\sqrt {{A_t}g}t$ are also plotted together with the linear fitting curves of the late-time stage, thus the spike and bubble growth rates by Method 5 can be extracted from the slopes of the fitting curves, yielding the values of 0.1135 and 0.0839, respectively. As summarized in Table I, we can observe that these statistical methods have different performances in predicting late-time growth rates of RTI, although they are totally equivalent in mathematics. Concretely, the spike prediction of Method 4 is much smaller than those of other statistical methods, and also the computed growth rates by Method 5 is slightly low. In general, the results of Methods 1, 2, and 3 approach to each other in predicting the growth rates of 3D single-mode RTI, and further combining the results of 2D single-mode case that Methods 1 and 5 are preferential, thus Method 1 would be recommended in the computations of the spike and bubble growth rates of single-mode RTI, which conforms to the report of multi-mode Rayleigh-Taylor turbulence~\cite{Cabot, Akula}.

\begin{table*}
\caption{\label{table1}The measured spike and bubble growth rates $\alpha_s$ and $\alpha_b$ using five statistical methods}
\begin{tabular}{cccc}
\hline
Statistical method ~~~~~~& Mathematical formula~~~~~~& $\alpha_s$ ~~~~~~& $\alpha_b$ \\
 \hline
Method 1  & $\alpha_{s,b}={\frac{{{\dot{h}}_{s,b}}^2}{4{A_t}gh}}$ ~~~& 0.1601 ~~~& 0.1008~~~  \\
Method 2  & ${\alpha_{s,b}}={\frac{h_{s,b}}{{A_t}gt^2}}$ ~~~& 0.2060 ~~~& 0.1127~~~  \\
Method 3  & ${\alpha_{s,b}}=={\frac{1}{2{A_t}} \frac{\partial{h_{s,b}}}{\partial{z}}}$~~~& 0.1790 ~~~& 0.1030~~~  \\
Method 4  & $\alpha_{s,b}={\frac{{{\ddot{h}}_{s,b}}}{2{A_t}g}}$ ~~~& 0.0489 ~~~& 0.1476~~~  \\
Method 5  & $h^{1/2}_{s,b}={({\alpha_{s,b}}{A_t}g)^{1/2}}t$~~~& 0.1135 ~~~& 0.0839~~~  \\
\hline
\end{tabular}
\end{table*}

\subsection{ Effect of the Atwood number}

\begin{figure}
\subfigure[]{\includegraphics[width=6cm]{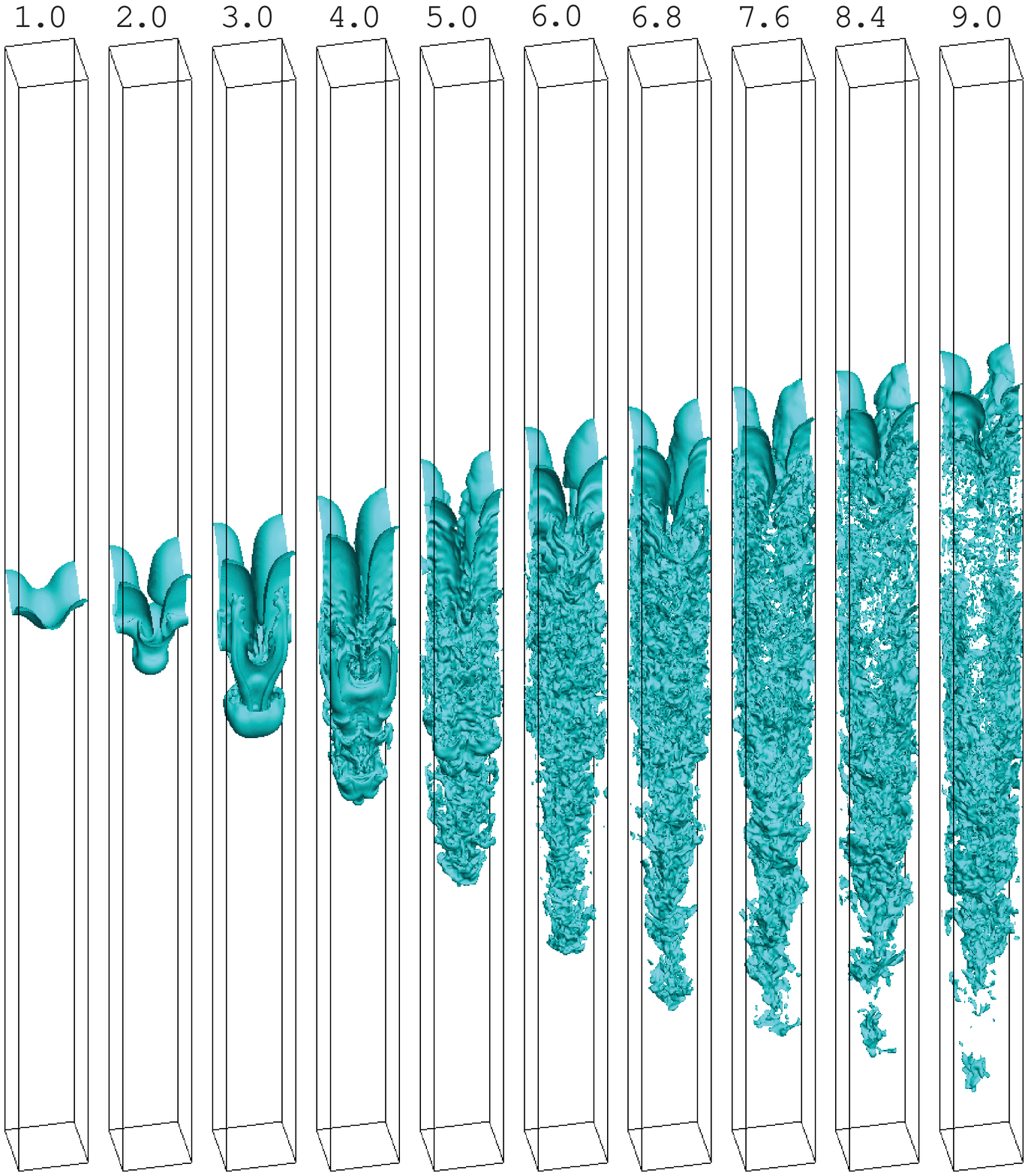}}
\subfigure[]{\includegraphics[width=6cm]{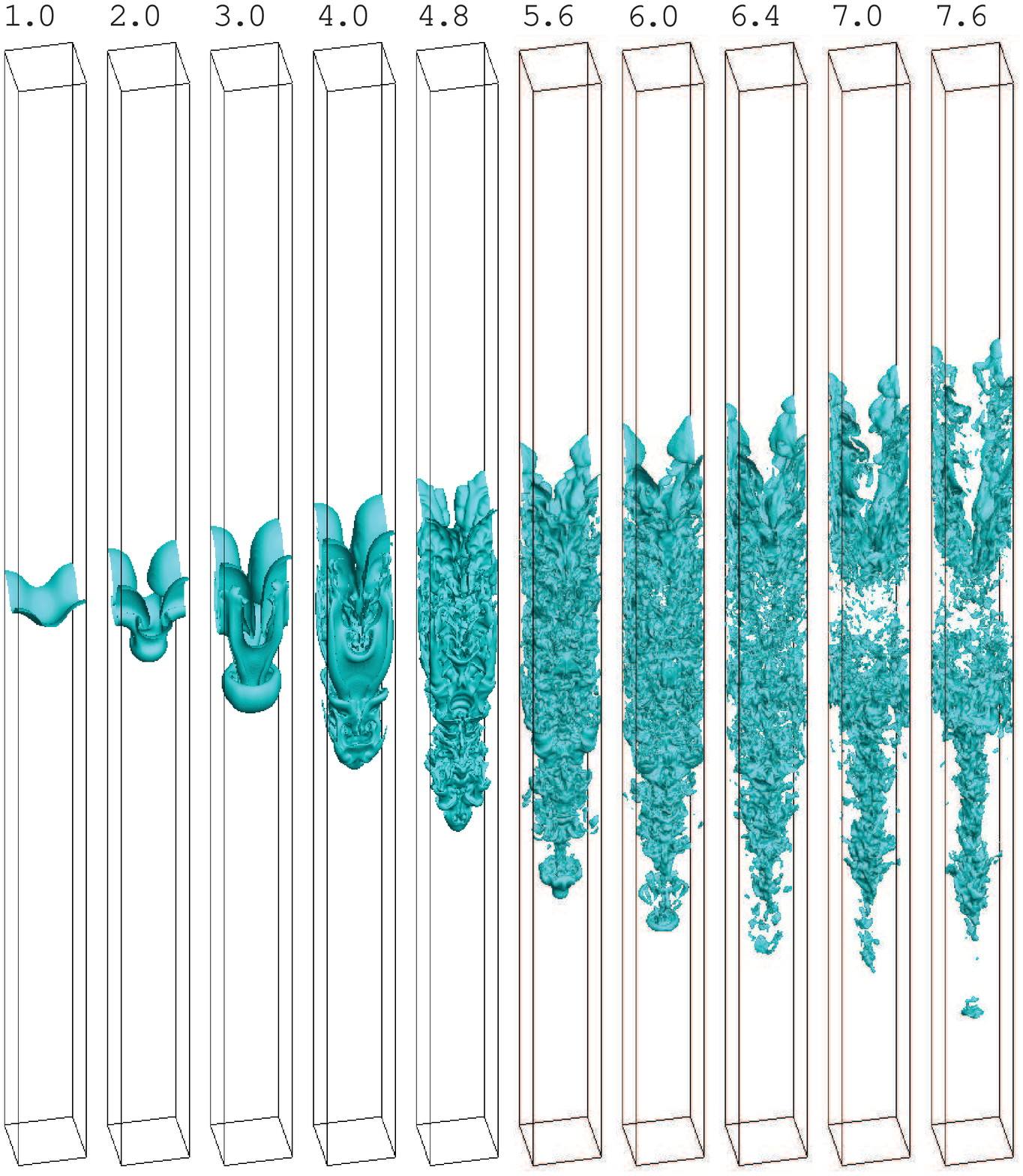}}
\subfigure[]{\includegraphics[width=6cm]{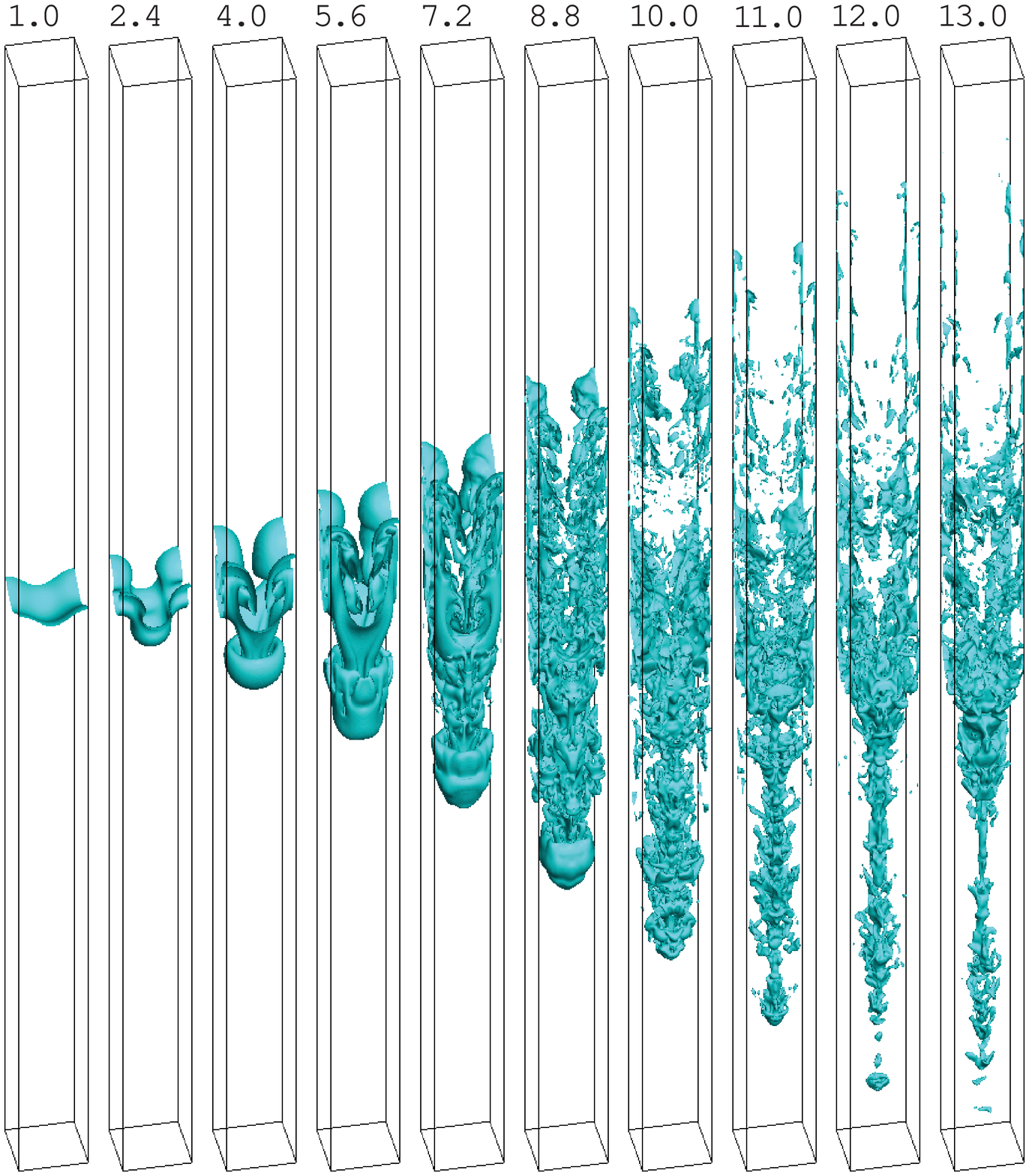}}
\subfigure[]{\includegraphics[width=6cm]{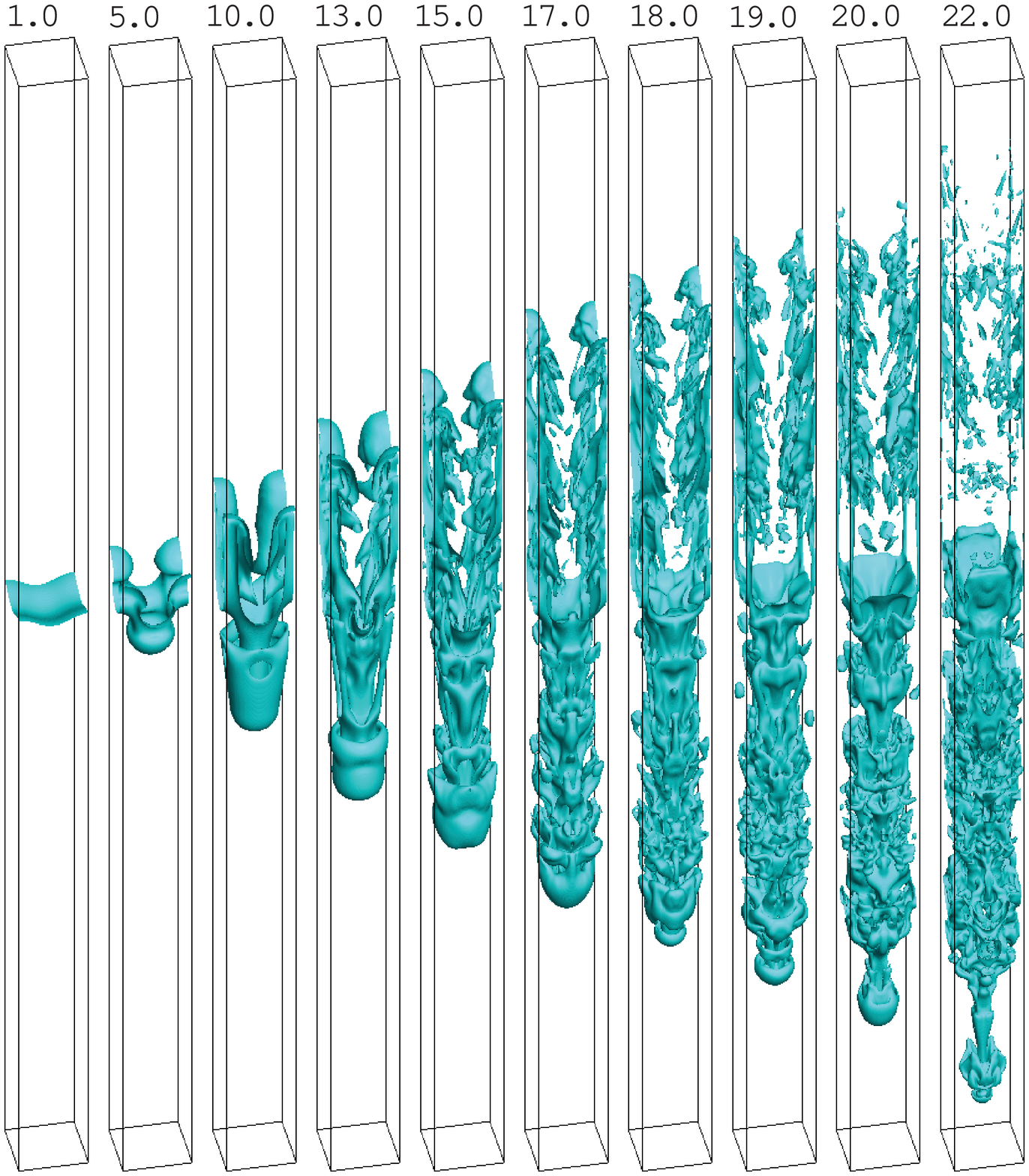}}
\caption{Time evolution of interfacial pattern in 3D immiscible RTI at various Atwood numbers, $Re=5000$: (a) $A_t=0.7$, (b) $A_t =0.6$, (c) $A_t=0.3$, (d) $A_t =0.1$.}
\end{figure}

\begin{figure}
\subfigure[]{\includegraphics[width=6cm]{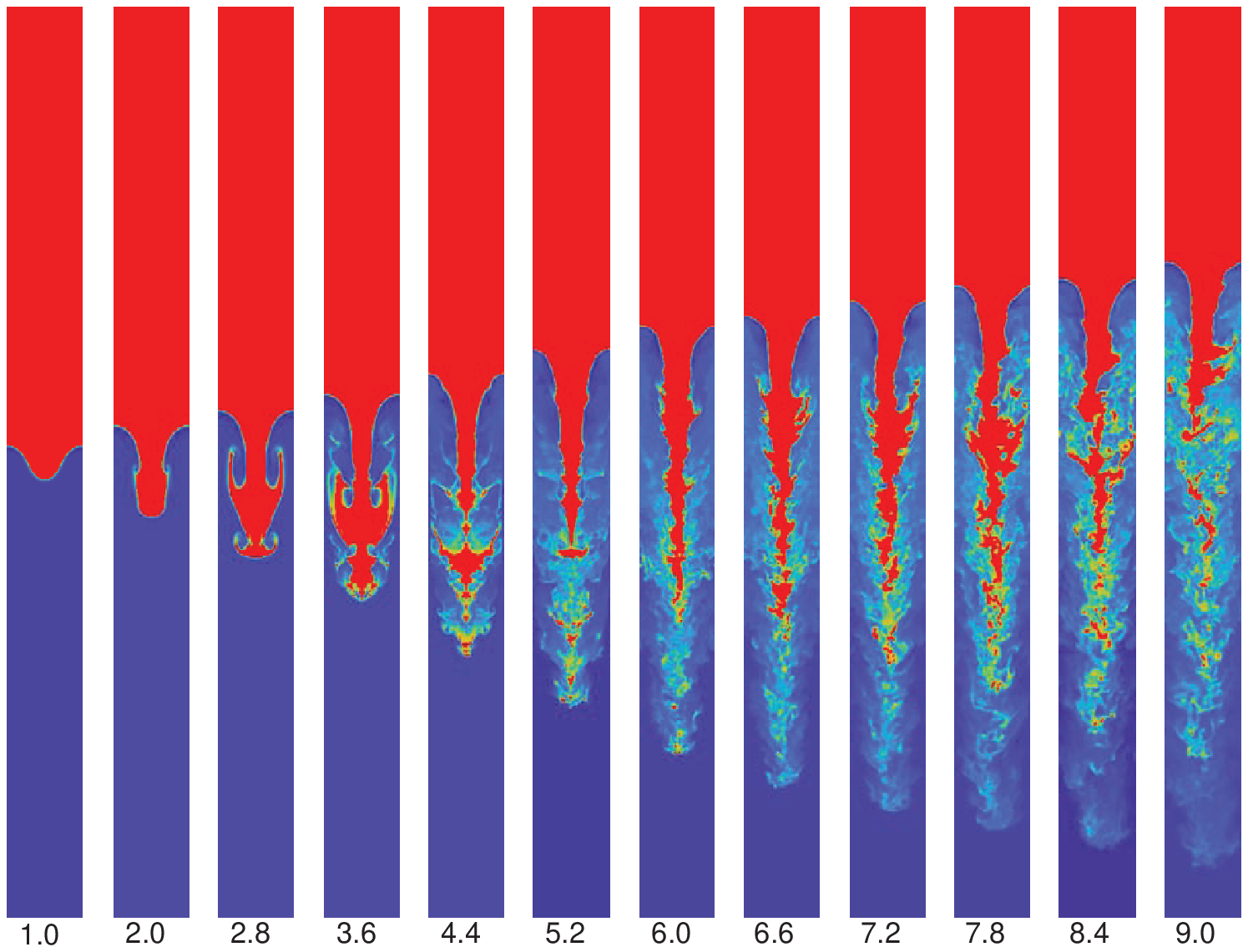}}
\subfigure[]{\includegraphics[width=6cm]{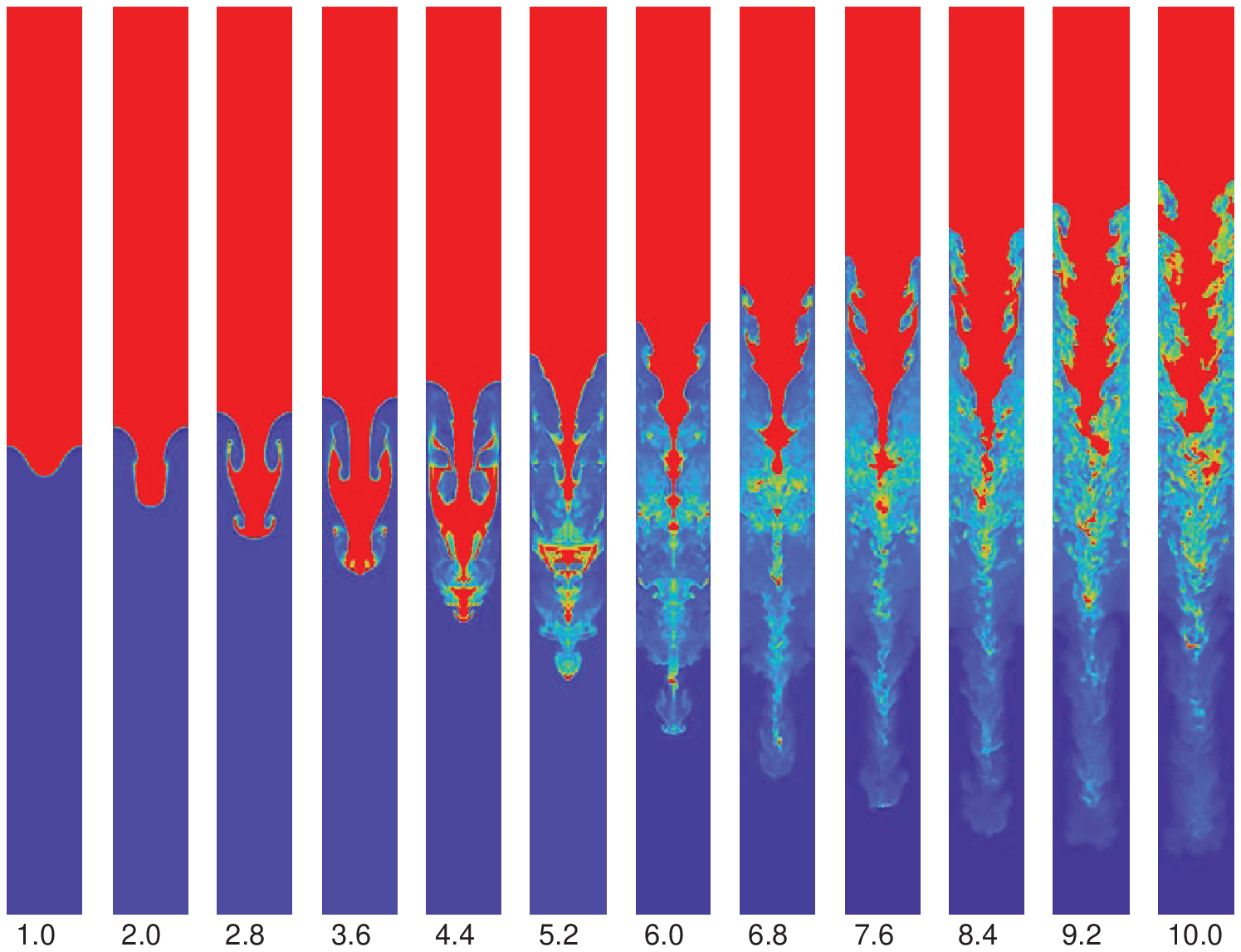}}
\subfigure[]{\includegraphics[width=6cm]{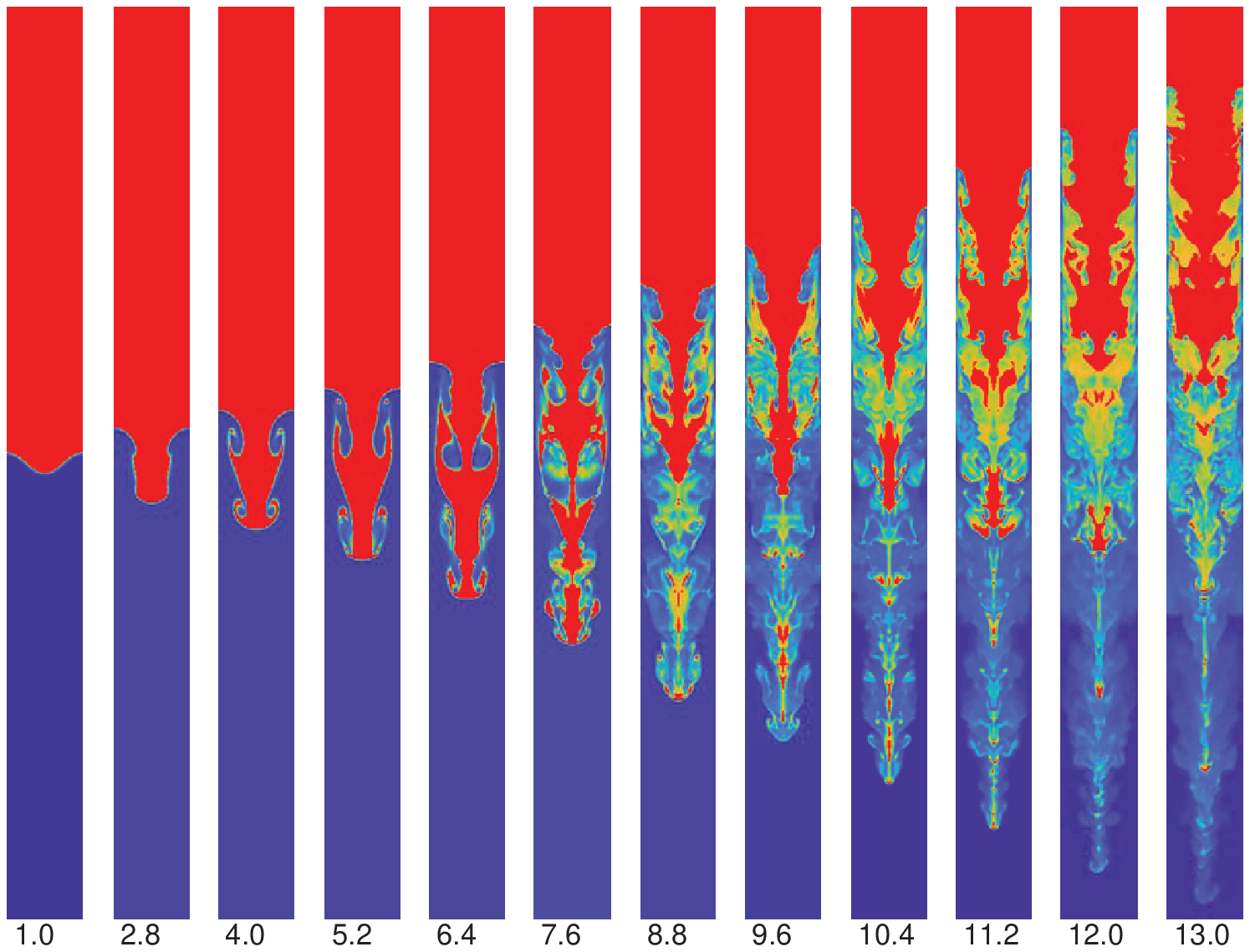}}
\subfigure[]{\includegraphics[width=6cm]{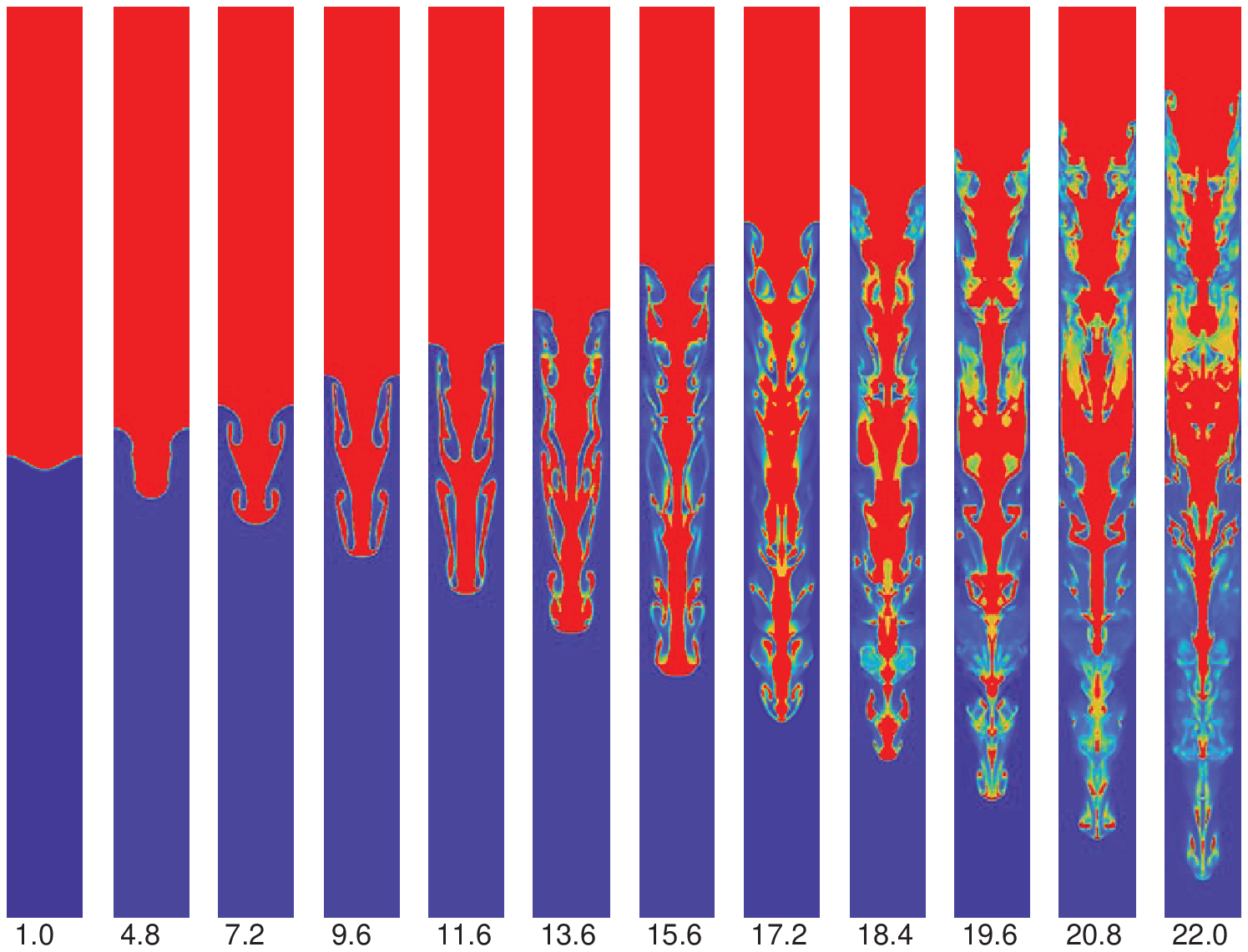}}
\caption{Time evolution of the density contour at the diagonal vertical plane, $Re=5000$: (a) $A_t =0.7$, (b) $A_t=0.6$, (c) $A_t=0.3$, (d) $A_t=0.1$.}
\end{figure}
In this subsection, the effect of the Atwood number on the late-time evolution of 3D single-mode RTI is investigated at a high Reynolds number of 5000. Figure 7 shows the snapshots of interfacial patterns in 3D single-mode RTI with different Atwood numbers. It can be observed that the heavy and light fluids regardless of the Atwood number penetrate into each other forming the spike and bubble at the initial stage, and then the spike rolls up at its end owing to the influence of Kelvin-Helmholtz instability followed by the generation of vortex structure. However, the vortex structure appears at later time for a smaller Atwood number. The vortex structure continues to develop in size for all cases and an interesting mushroom-like shape can be visible in the system. After that, the secondary vortices can be generated and even the multiple-layer roll-up behaviours occur due to the increasing strengthes of the Kelvin-Helmholtz vortices. Finally, the interfacial instability is aggravated at the late time, facilitating the completely turbulent state of mixing layer. The close inspection of interfacial dynamics in 3D single-mode RTI shows that the symmetry of interface is broken for a Atwood number higher than 0.3, while it can still preserve the symmetry with respect to the middle line for a low Atwood number less than 0.3. This result conforms to the finding of high-resolution direct numerical simulation of single-mode RTI with a low Atwood number that the symmetry of interfacial pattern can be persistently maintained~\cite{Wei}. To observe the evolution of the interface more clearly, we also presented the density image at the diagonal vertical plane with above Atwood numbers in Fig. 8. A unique feature of two pairs of counter-rotating vortices can be observed for all Atwood numbers, while it emerges at later time as the Atwood number is decreased. During the chaotic mixing stage, the interaction between fluids becomes intensive and the interface could undergo a dramatic deformation. The lines of symmetry within the bubble and spike of 3D single-mode RTI are clearly preserved at a low Atwood number of 0.1, but are broken in the cases of higher Atwood numbers.
\begin{figure}
\subfigure[]{\includegraphics[width=3.0in,height=2.5in]{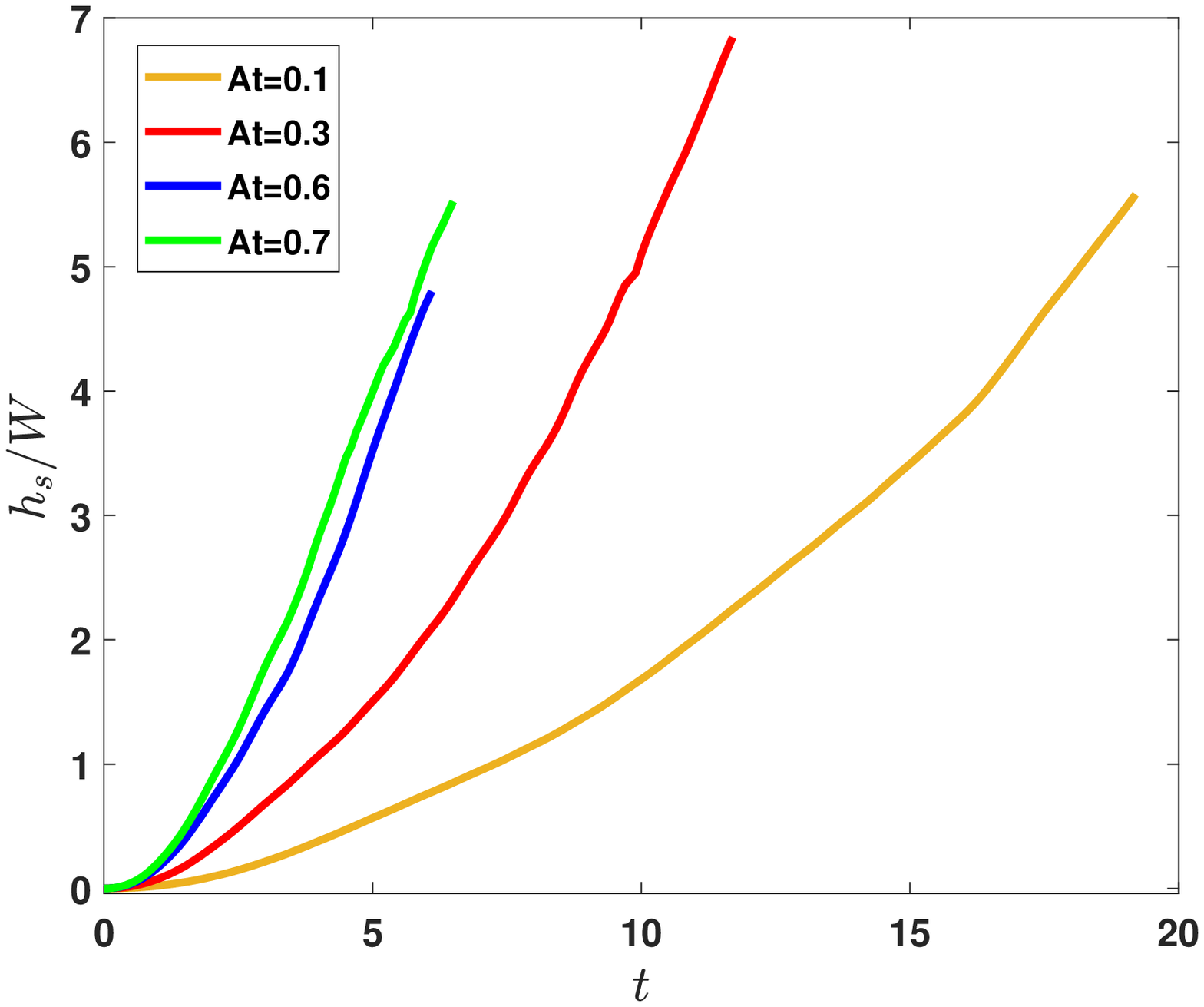}}
\subfigure[]{\includegraphics[width=3.0in,height=2.5in]{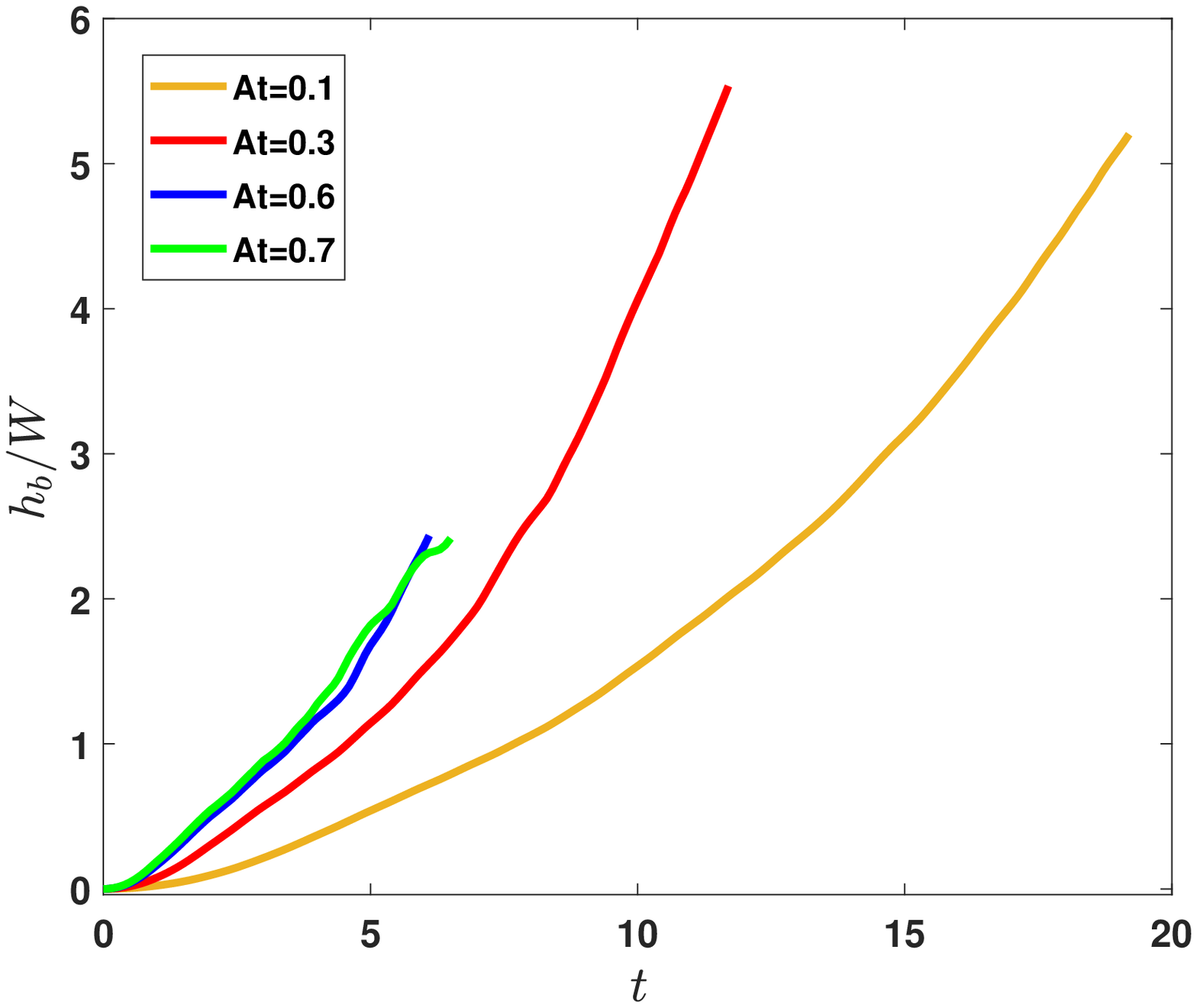}}
\caption{Effect of the Atwood number on the (a) normalized spike amplitude and (b) normalized bubble amplitude in 3D single-mode RTI at $Re=5000$.}
\end{figure}

\begin{figure}
\centering
\subfigure[]{\includegraphics[width=3.0in,height=2.5in]{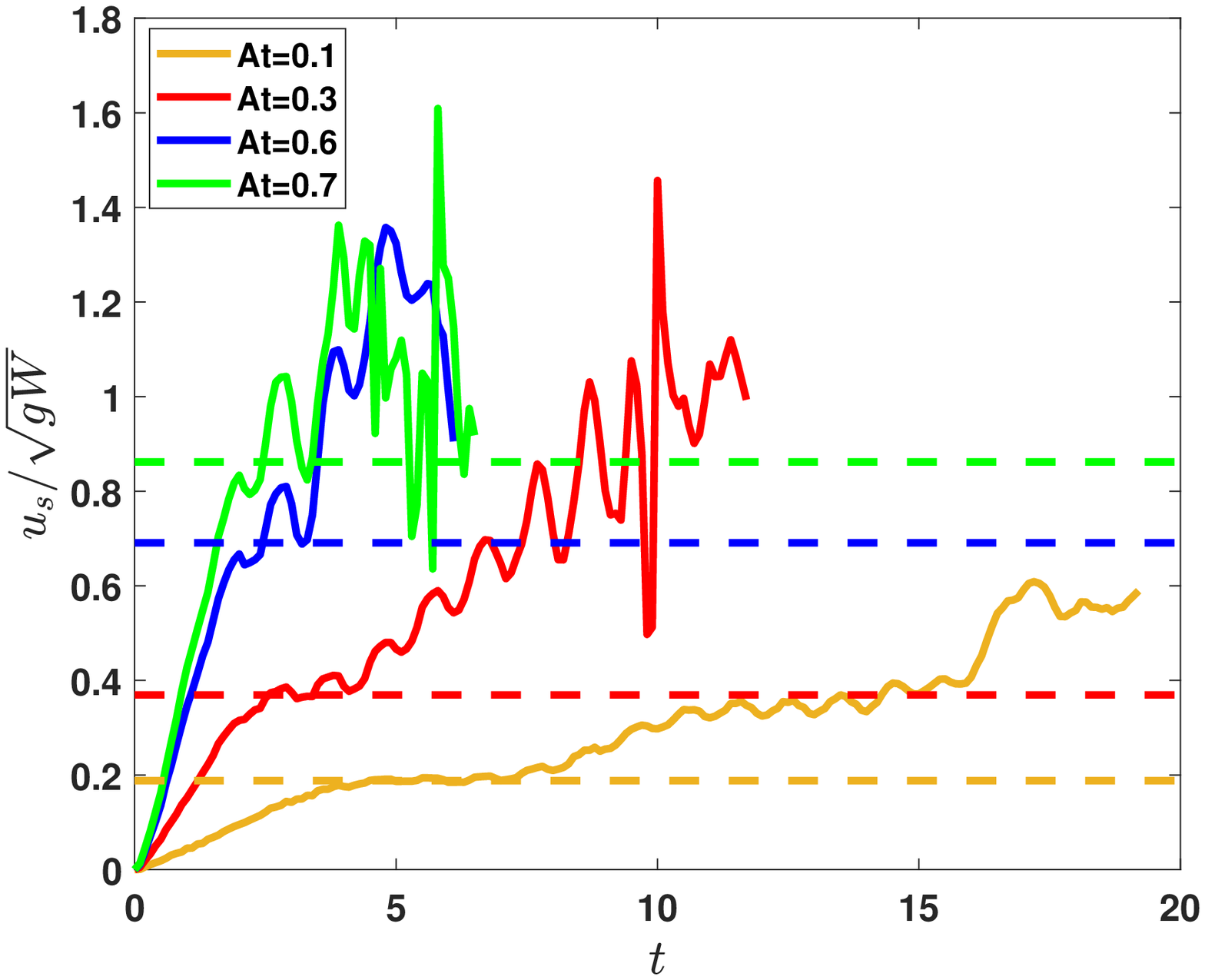}}
\subfigure[]{\includegraphics[width=3.0in,height=2.5in]{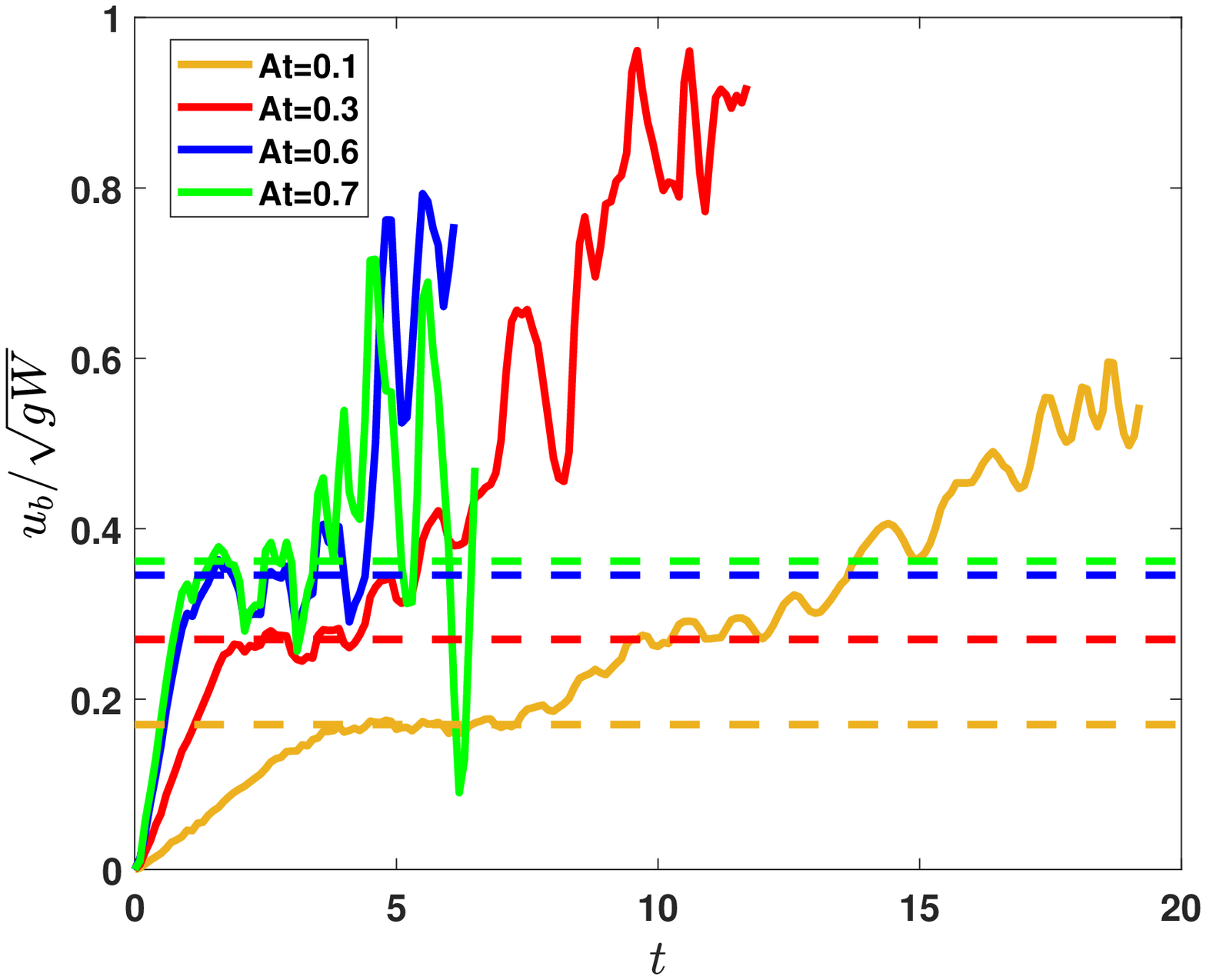}}
\caption{Effect of the Atwood number on the (a) normalized spike velocity and (b) normalized bubble velocity in 3D single-mode RTI with $Re=5000$. The dotted lines mark the analytical solutions of the potential flow model proposed by Goncharov\cite{Goncharov} at different Atwood numbers.}
\end{figure}
Figure 9 shows the time evolutions of the normalized spike and bubble amplitudes with different Atwood numbers. It can be found from this figure that the spike and bubble amplitudes increase with time, and achieve greater values at a higher Atwood number, implying the disturbance of the single-mode instability grows faster with the increase of the Atwood number. In addition, the comparison between the curves of the spike and bubble amplitudes showed the asymmetric developments for the spike and bubble fronts, which become more significant as the Atwood number is increased. Figure 10 depicts the time variations of the normalized spike and bubble speeds under the corresponding Atwood numbers. For all Atwood numbers, we can observe that the 3D single-mode RTI at a high Reynolds number also experiences four different development stages: the linear growth, saturated velocity, reacceleration and turbulent mixing stages. Following the initial stage, the 3D single-mode instability enters into the second saturated velocity stage characterizing with approximately constant velocities, and the numerical predictions for the spike and bubble velocities are in good agreements with the analytical solutions of the potential flow model, both of which increase with the Atwood number. In addition, we can learn that the saturated stage would be reached at the earlier time for a higher Atwood number, and also the duration of the saturated stage obviously decreases with the Atwood number. This is because that a greater buoyancy according to Eq. (32) is achieved for a larger Atwood number, making the spike and bubble grow more quickly and also a faster increasing in the strength of the Kelvin-Helmholtz vortices. In this case, the spike and bubble are easily accelerated to exceed than their asymptotic values and enter into the reacceleration stage. In the turbulent mixing stage, the bubble and spike are accelerated and decelerated repeatedly, exhibiting the fluctuated behaviours for all Atwood numbers. However, the range of the fluctuation seems to be reduced at a smaller Atwood number.

\begin{figure}
\subfigure{\includegraphics[width=4.2in,height=3.4in]{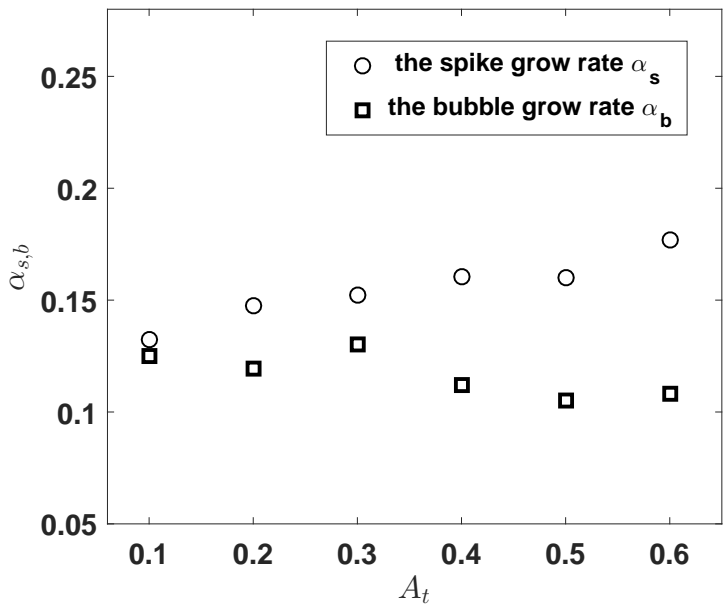}}
\caption{The late-time growth coefficients ${\alpha_{s,b}}$ of the bubble and spike at varies Atwood numbers}
\end{figure}

To quantitatively describe the nature of late-time growth of 3D single-mode RTI, we measured the spike and bubble growth rates with various Atwood numbers using the statistical approach given by Eq. (35) and presented the corresponding curves in Fig. 11. It can be found that for a fixed Atwood number, the spike growth rate is greater than that of the bubble implying the asymmetric development between the spike and bubble, and this trend becomes more significant as the fluid Atwood number increases. In addition, we can observe that the spike growth rate shows an overall increase with the Atwood number and the bubble growth rate is basically not affected by the Atwood number approaching to a constant around 0.1. In a comparison with the 2D report~\cite{Liang1}, we can further find that the growth rates of 3D single-mode RTI for both the spike and bubble are obviously larger than those of 2D example for a same Atwood number.

\section{conclusions}~\label{sec:summary}
In this paper, we used an improved mesoscopic lattice Boltzmann model to investigate the late-time dynamics of 3D single-mode RTI and mainly examined the effects of the Reynolds number and Atwood number on the growth of the spike and bubble fronts. The numerical experiments indicate that the development of 3D single-mode RTI at a high Reynolds number can be summarized into four stages, including linear growth, saturated velocity growth, reacceleration and turbulent mixing stages. We observe that the spike and bubble at the second stage grow with approximately constant speeds, and their values are consistent with the analytical solutions of the potential flow theory~\cite{Goncharov}. In addition, the duration of the saturated velocity stage obviously decreases with the Atwood number. At the late time of evolution, the phase interface would undergo a large deformation and even a chaotic breakup, promoting the fully turbulent mixing of fluids. The lines of symmetry within the bubble and spike of 3D single-mode RTI can be always preserved for the case of a low Atwood number, but are clearly broken at a high Atwood number. The spike and bubble late-time velocities fluctuate with time, exhibiting an averaging quadratic growth law in the turbulent mixing stage. To quantitatively reveal the late-time growth law of 3D single-mode RTI, five statistical approaches for calculating the growth rate are discussed and a preferential one is recommended. It can be found that the spike growth rate increases with the Atwood number, while the bubble growth rate is little affected by the change of the Atwood number, approaching a constant of around 0.1. When Reynolds number is low, the later stages cannot be reached sequentially and the evolution of the phase interface presents a laminar flow state.

\section*{Acknowledgments}
This work is financially supported by the National Natural Science Foundation of China (Grant Nos. 11972142, 51976128).


\end{document}